\documentclass[12pt]{iopart}

\usepackage[T1]{fontenc}
\usepackage[latin1]{inputenc}
\usepackage{graphicx}
\usepackage{amssymb}
\usepackage{esint}

\makeatletter

\usepackage{epsfig}

\usepackage{dcolumn}

\usepackage{bm}

\usepackage{bbm}

\usepackage{wasysym}

\usepackage{marvosym}

\usepackage{hyperref}

\newcommand{\ud}{\mathrm{d}}
\newcommand{\f}{\varphi}
\newlength{\textwidthm}

\makeatother

\begin{document}

\title{Dirac electrons in  graphene-based quantum wires and quantum dots}

\author{N.~M.~R.~Peres$^1$, J.~N.~B.~Rodrigues$^2$, T. Stauber$^1$, and J.~M.~B. Lopes dos Santos$^2$}

\address{$^{1}$Centro de F\'{\i}sica e Departamento de F\'{\i}sica, Universidade do Minho, P-4710-057, Braga, Portugal}

\address{$^{2}$ CFP and Departamento de F\'{\i}sica, Faculdade de Ci\^{e}ncias
Universidade do Porto, P-4169-007 Porto, Portugal}

\date{\today}

\begin{abstract}
In this paper we analyse the electronic properties of Dirac electrons
in finite-size ribbons and in circular and hexagonal quantum dots. We
show that due to the formation of sub-bands in the ribbons it is
possible to spatially localise some of the electronic modes using a
$p-n-p$ junction.  We also show that scattering, by an
infinitely-massive wall, of confined Dirac electrons in a narrow
channel, induces mode mixing, giving a qualitative reason for the fact
that an analytical solution to the spectrum of Dirac electrons
confined in a square box has not been found, yet. A first attempt to the
solution of the square billiard is presented. We find that only the
trivial case $k=0$ has a solution that does not require the existence
of evanescent modes.  We also study the spectrum of quantum dots of
graphene in a perpendicular magnetic field.  This problem is studied
in the Dirac approximation, and its solution requires a numerical
method whose details are given.  The formation of Landau levels in the
dot is discussed. The inclusion of the Coulomb interaction among the
electrons is considered at the self-consistent Hartree level, taking
into account the interaction with an image-charge density necessary to
keep the back-gate electrode at zero-potential. The effect of a radial
confining potential is discussed.  The density of states of circular
and hexagonal quantum dots, described by the full tight-binding model,
is studied using the Lanczos algorithm. This is necessary to access
the detailed shape of the density of states close to the Dirac point
when one studies large systems.  Our study reveals that zero energy
edge states are also present in graphene quantum dots. Our results
are relevant for experimental research in graphene nanostructures.
The style of writing is pedagogical, hoping that new-comers to the
subject can find this paper a good starting point for their research.
\end{abstract}

\pacs{73.21.Hb, 73.21.La, 73.23.-b, 73.43.-f, 73.63.Kv, 73.63.Nm}

\section{Introduction}

Graphene was discovered in 2004 at the Centre for Mesoscopic and
Nanotechnology of the University of Manchester, U.K., directed by
A. K. Geim \cite{novo1,pnas}. Previously, graphene was known only as
an intrinsic part of three-dimensional systems: as individual atomic
planes within graphite or its intercalated compounds and as the top few
layers in epitaxially grown films \cite{japs}. In certain cases, it
was possible to even grow graphene monolayers on top of metallic
substrates and silicon carbide \cite{japs}. However, coupling with the
substrate did now allow studies of electronic, optical, mechanical,
thermal and other properties of graphene, which all became possible
after individual graphene layers were isolated. 
  There are by now a number of review papers on
graphene available in the literature, both of qualitative
\cite{castro,Nov07,katsnelson,Geim07,ScA} and quantitative
\cite{rmp,rmpBeenakker} nature.

The original method of graphene isolation is based on
micro-mechanical cleavage of graphite surface -- the so called {\it
scotch tape method}. This method, however, has a low yield of graphene
micro-crystallites. Recently, a new method \cite{Nature,NL_stiffness},
based on liquid-phase exfoliation of graphite has proved to produce a
large yield of graphene micro-crystallites, with large surface areas.
A chemical approach to graphene production has also been achieved
using exfoliation-reintercalation-expansion of graphite \cite{dai08}.

It is by now well known that graphene is a one-atom thick sheet of carbon
atoms, arranged in a honeycomb (hexagonal) lattice, having therefore
two carbon atoms per unit cell. The material can be considered the
ultimate thin film.  In a way, this material was the missing allotrope
of pure carbon materials, after the discovery of graphite
\cite{book_carbon,chung02}, diamond \cite{book_diamond}, fullerenes,
and carbon nanotubes \cite{book_carbon}.  In fact we can think of
graphene as being the raw material from which all other allotrope's of
carbon can be made \cite{castro,Nov07}.

Although the Manchester team produced two-dimensional
micro-crystallites of other materials \cite{pnas}, graphene attracted
a wide attention \cite{scientometrics} from the community due to its
unexpected properties, both associated with fundamental and applied
research.

Graphene has a number of fascinating properties. The stiffness of
graphene has been proved to be so large, having a Young modulus $E\simeq
1.0$ TPa, that  makes it the strongest material-stiffness ever measured
\cite{NL_stiffness,S_stiffness}.  In addition, the material has high
thermal conductivity \cite{lau}, is chemically stable and almost
impermeable to gases \cite{ballon}, can withstand large current
densities \cite{novo1}, has ballistic transport over sub-micron scales
with very high mobilities in its suspended form ($\mu\simeq$ 200.000
cm$^2\cdot$V$^{-1}\cdot$s$^{-1}$) \cite{novo1,giantmob,giantkim}, and shows am-bipolar
behaviour \cite{novo1}. Ballistic transport is in general associated
with the observation of conductance quantisation in narrow channels
\cite{peresG}, which have recently been observed \cite{avouris}.  The
above properties and its two-dimensional nature makes graphene a
promising candidate for nano-electronic applications.

From the point of view of physical characterisation, we are interested
in the mechanical properties of graphene, its electronic spectrum, its
transport properties of heat, charge and spin, and its optical
properties. The high stiffness of the material is responsible for the
micro-crystallites to keep their planar form over time, without
rolling up, even when graphene is held fixed by just one of its ends
\cite{NL_stiffness}. Interestingly, graphene is now being used,
by combining electrostatic deposition methods and the chemical nature
of the surrounding atmosphere, to produce rolled up nanotubes with
controlled dimensions and chiralities \cite{graphene_nanotube}.  The
electronic spectrum of graphene and of graphene bilayer has been
measured by angle resolved photoemission spectroscopy (ARPES)
\cite{rotemberg1,rotemberg2,lanzara1,lanzara2}, fitting well with
tight-binding calculations using a first nearest neighbour hopping
$t\simeq 3$ eV and a second nearest neighbour hopping 
$t'\simeq 0.13$ eV \cite{Deacon}.

The transport of heat has been measured experimentally and studied
theoretically in few papers
\cite{lau,Heat_peres,Heat_stauber,Lofwander,Trushin}, and more
research is needed to fully understand its properties, especially
since the irradiation of graphene with laser light has been found to
heat up locally the system \cite{lau}. The transport
properties of graphene on top of silicon oxide have been extensively
studied, but in a suspended geometry the few available experimental and
theoretical studies are still recent
\cite{giantkim,Eva_Andrei,sarma,stauber_sus}. Of particular interest is the
contribution of phonons to the transport properties of graphene.
Although in the beginning of graphene research phonons were considered
not important, recent experimental results show that this is, in fact,
not the case \cite{giantmob,rotemberg3}.

Of particular interest is the finite conductivity of graphene at the
Dirac point, $\sigma_D$, whose measured value is in contradiction to
the naive single particle theory, which predicts either infinite or
zero resistance. The value of $\sigma_D$ is of the order of
\begin{equation}
 \sigma_D \simeq \lambda 4\frac {e^2}{h}\,,
\label{sD}
\end{equation}
with $\lambda$ a number unity order.  We emphasise that Eq. (\ref{sD})
is the value for the conductivity of the material at the Dirac point,
and not the conductance of a narrow channel.  On the other hand, there
is a discrepancy between the more elaborated theoretical descriptions
of $\sigma_D$ and the experimental measured values, since the theory
predicts the value $\sigma_D^{theor.}=\sigma_D/\pi$
\cite{peres_prb,beenakker06,katsnelson06,ziegler07}, and most of the
experiments measure a value given by Eq. (\ref{sD}). Adding to the
problem, two experimental groups reported measurements of the
conductivity of graphene consistent with the theoretical calculation
\cite{laupi,morpurgo}. In this context, it should be stressed that
whereas the result of $\sigma_D$ obtained in Ref. \cite{peres_prb}
comes about due to an increase of the density of states due to
disorder (albeit small) at the Dirac point, the value for $\sigma_D$
computed in Refs. \cite{beenakker06,katsnelson06} is based on the
existence of evanescent waves in clean graphene ribbons with large
aspect ratio $W/L$ ($W$ is the width and $L$ in the length of the
ribbon). The recent experiments \cite{laupi,morpurgo} seem to confirm
this latter view of the problem, since the value $\sigma_D^{theor.}$
is only measured in the regime $W/L\gg 1$. The transport of
spin in graphene was studied experimentally in few publications
\cite{ieee,apl_spin,nature_spin,spin1}, and much work remains to be
done.

The optical properties of graphene and of bilayer graphene have only
recently been studied experimentally \cite{basov1,basov2}, in
contrasts to the corresponding theoretical studies. The first
theoretical study of the graphene's optical absorption was done by
Peres {\it et al.}  \cite{peres_prb}, followed by several studies by
Gusynin {\it et al.} and reviewed in \cite{sharapovIJMPB}. The most
relevant aspect was that the infrared conductivity of graphene, for
photon energies larger than twice the chemical potential, has a
universal value given by \cite{peres_prb,sharapovIJMPB,Stauber_prb_08}
\begin{equation}
 \sigma_0=\frac \pi 2 \frac {e^2}h\,.
\label{s0}
\end{equation}
Also for the bilayer, the theoretical studies preceded the
experimental measurements \cite{Nilsson06,Falko07,Nicol08}. Due to the
existence of four energy bands in bilayer graphene, its optical
spectrum has more structure than the corresponding single layer one.

It was experimentally found that for photon energies in the visible
range \cite{Nair} Eq. (\ref{s0}) also holds within less than 10\%
difference \cite{Stauber_Geim_prb_08}. This result makes graphene the
first conductor with light transmissivity, in the frequency range
from infrared to the ultraviolet, as high as

\begin{equation}
 T\simeq 1-\pi\alpha\sim 98\%\,,
\end{equation}
with $\alpha$ the fine structure constant, in a frequency range that
extends from the infrared to the ultraviolet. This makes obvious that
graphene can be used as a transparent metallic electrode, having found
applications in solar cell prototypes \cite{Wang,Wu} and in gateable
displays \cite{blake}. The same conclusions are obtained from studying
the optical conductivity of graphite \cite{Kuzmenko}. The transparency
of graphene has an obvious advantage over the more traditional
materials, used in the solar cell industry, indium tin oxide (ITO) and
fluorine tin oxide (FTO), which have a very low transmission of light
for wave-lengths smaller than 1500 nm; in the visible range the
transparency of these two materials is larger than $\sim
85\%$. Furthermore, these traditional materials have a set of
additional problems \cite{Wang,blake}, such as chemical instability,
which are not shared by graphene. On the other hand, ITO and FTO have
a low resistivity ($\sim 5\Omega\cdot$m), a figure that graphene cannot
match, if one leaves aside the possibility of graphene films deposit
from solution.

Finally, the interaction of graphene with single molecules allows to
use graphene as a detector of faint amounts of molecules
\cite{sensors} and to enhance, in a dramatic way, the sensibility of
ordinary transmission electron microscopes \cite{Nature,Zettl_tem}, allowing
the observation of adsorbates, such as atomic hydrogen and oxygen,
which can be seen as if they were suspended in free space.

Many of the above properties are expected to be present in graphene
nanoribbons.  On the other hand, aspects related to the quantum
confinement of electrons in graphene, both considering confinement in
one- (quantum wire) or two- (quantum dot) dimensions, is expected to
bring new interesting phenomena. The present state of the art of
material manipulation technologies does not allow to produce
structures, on a top-down approach, smaller than 10 nm. Nevertheless,
many aspects associated with the quantum confinement of electrons in
graphene either due to narrow constrictions or due to the formation of
quantum dots have already been investigated experimentally
\cite{freitag,dairibbon,geimdots,ensslina,ensslinb} and theoretically
\cite{Efetov,Sols,Wunsch,KatsnelsonPaco}. One important consequence of
quantum confinement is the appearance of an energy gap in the
electronic spectrum of graphene, an important characteristic if
graphene is to be used as a material to build nano-transistors
\cite{daiprl}.  In this paper we will address several properties of
confined Dirac electrons, by considering both nano-wires and quantum
dots of graphene. In doing this we use both the continuous Dirac
approximation and the tight-binding description, choosing which one is
more appropriate to the given problem.

\section{The tight-binding model and the Dirac approximation}

\begin{figure}[t]
\begin{center}
\includegraphics*[angle=0,width=8cm]{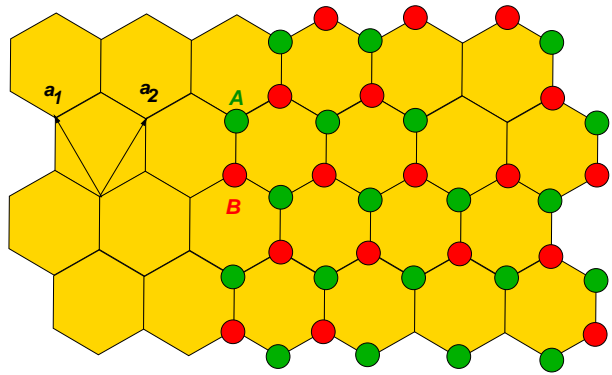}
\caption{(colour on-line) An honeycomb ribbon, with zigzag edges (top and bottom)
and arm-chair edges (vertical ones), with the carbon atoms belonging
to the sub-lattices $A$ and $B$ clearly differentiated from each other (with several carbon atoms
represented).
The lattice unit vectors $\bm a_1$ and $\bm a_2$ are also shown. 
\label{fig1}}
\end{center}
\end{figure}

The experimental work cited in the introduction constitutes a vast
evidence that the low-energy theory of electrons in graphene is
described by the two-dimensional Dirac equation, which is obtained as
an $ \bm k\cdot \bm p$ expansion around the Dirac points in momentum
space \cite{rmp,mele}. In Figure \ref{fig1} we represent a finite-size
ribbon of an hexagonal lattice. Two features are of importance: the
first is that the lattice is not a Bravais lattice, being instead made
of two inter-penetrating triangular lattices, giving rise to two
geometrically nonequivalent carbon atoms, termed $A$ and $B$; the
second is that there are two different types of edges present --
zigzag and arm-chair edges. These types of edges play a different role
in the physics of the ribbon. In particular, note that the zigzag
edges are constituted by a single type of atom, $B$ on top and $A$ at
the bottom (in the case of this figure).  As is shown in
Fig. \ref{fig1}, we can choose the direct lattice vectors to be the
following:
\begin{eqnarray}
{\bm a}_{1} & = & \frac{a}{2} ( - 1 , \sqrt{3} ), \label{eq:BasisVectorsHoneycomb1} \\
{\bm a}_{2} & = & \frac{a}{2} ( 1 , \sqrt{3} ), \label{eq:BasisVectorsHoneycomb2}
\end{eqnarray}
where $a=2.46$ \AA\,  is the lattice vector length. As a consequence, the reciprocal lattice vectors are
\begin{eqnarray}
{\bm b}_{1} & = & \frac{2 \pi}{a} ( - 1 , \frac{1}{\sqrt{3}} ), \\
{\bm b}_{2} & = & \frac{2 \pi}{a} ( 1 , \frac{1}{\sqrt{3}} ).
\end{eqnarray}
The so called Dirac points in the honeycomb Brillouin zone are conveniently chosen to be
\begin{eqnarray}
{\bm K} & = & \frac{4 \pi}{3 a} (1 , 0 ), \\
{\bm K}' & = & \frac{4 \pi}{3 a} (- 1 , 0 )\,.
\end{eqnarray}
If one considers only the hopping process $t$ (first nearest neighbour hopping), the tight-binding Hamiltonian is very easily written as ($N_c$ is the number of unit cells in the solid)
\begin{equation}
 H=-t\sum_{i,\sigma}^{N_c}(a^\dag_{i,\sigma}b_{i,\sigma} + {\rm H.\,c.})\,,
\label{Htb}
\end{equation}
where $a^\dag_{i,\sigma}$ creates an electron with spin projection $\sigma$ in the $\pi$-orbital of the carbon atom of the sub-lattice
$A$, and of the unit cell $i$; a similar definition holds for $b^\dag_{i,\sigma}$.
The exact diagonalisation of this problem
is straightforward, leading to 
\begin{eqnarray}
E_{\pm} = \pm t \sqrt{3 + 2 \cos (a k_{x}) + 4 \cos \left(\frac{a}{2} k_{x}\right) \cos \left(\frac{a \sqrt{3}}{2} k_{y}\right)}\,. \nonumber
\end{eqnarray}
We can expand this relation near the Dirac points, obtaining ($\boldsymbol{k} =\boldsymbol{q} + \boldsymbol{K}$)
\begin{eqnarray}
E_{\pm} \simeq \pm v_{F} | \vec{q} |\,,
\end{eqnarray}
which is a massless Dirac-like linear dispersion relation, where the velocity of light is substituted by $v_{F} = \frac{a \sqrt{3}}{2 \hbar} t \simeq 10^{6}$ m/s, the Fermi velocity. To obtain the effective Hamiltonian obeyed by the electrons near the
Dirac points we write the matrix Hamiltonian in momentum space as
\begin{equation}
 H_{\boldsymbol{k}} = - t \left ( \begin{array}{ccc} 0 & s_{\boldsymbol{k}} \\ s_{\boldsymbol{k}}^{*} & 0 \end{array} \right ),
\end{equation}
with $s_{\boldsymbol{k}}$ given by ($\boldsymbol{k} =\boldsymbol{q} + \boldsymbol{K}$)
\begin{eqnarray}
s_{\boldsymbol{k}} &=& 1 + e^{i \boldsymbol{K} \cdot \boldsymbol{a}_{1}} e^{i \boldsymbol{q} \cdot \boldsymbol{a}_{1}} + e^{i \boldsymbol{K} \cdot \boldsymbol{a}_{2}} e^{i \boldsymbol{q} \cdot \boldsymbol{a}_{2}} \ \nonumber \\
&\simeq&  - \frac{a \sqrt{3}}{2} ( q_{x} + i q_{y} ), \nonumber
\end{eqnarray}
leading to the effective Hamiltonian (one valid near $\boldsymbol{K}$ and the other near $\boldsymbol{K'}$ )
\begin{eqnarray}
\label{HK}
H_{\boldsymbol{K}} (\boldsymbol{q}) &=& v_{F} \boldsymbol{\sigma}^{*} \cdot \boldsymbol{q} \\ 
\label{HKP}
H_{\boldsymbol{K'}} (\boldsymbol{q}) &=& v_{F} \boldsymbol{\sigma} \cdot \boldsymbol{q}\,,
\end{eqnarray}
with  $\boldsymbol{\sigma} = ( \sigma_{x} , \sigma_{y} )$ and $\boldsymbol{\sigma}^{*} = ( \sigma_{x} , - \sigma_{y} )$.
The Hamiltonian (\ref{HK}), or alternatively (\ref{HKP}), will be the starting point of our discussion.
From Eqs. (\ref{HK}) and (\ref{HKP}) we can write the second-quantised Hamiltonian for electrons in graphene
\begin{eqnarray}
H \simeq - i v_{F} \int \ud x \ud y \Big( \hat{\Psi}_{1}^{\dagger} (\boldsymbol{r}) \bm\sigma \cdot \bm\nabla \hat{\Psi}_{1} (\boldsymbol{r}) + \hat{\Psi}_{2}^{\dagger} (\boldsymbol{r}) \bm\sigma^{*} \cdot \bm\nabla \hat{\Psi}_{2} (\boldsymbol{r}) \Big),
\end{eqnarray}
where $\hat{\Psi}_{i}^{\dagger} = (a_{i}^{\dagger}, b_{i}^{\dagger})$ (for $i = 1, 2$).

Let us assume that it is possible to create a potential such that a term of the form
\begin{equation}
 V=
\sum_{i,\sigma}^{N_c} v_{F}^{2} m ( a_{i,\sigma}^{\dagger}a_{i,\sigma} - 
b_{i,\sigma}^{\dagger} b_{i,\sigma}), 
\label{eq:TBMassTerm}
\end{equation}
is added to the Hamiltonian (\ref{Htb}). In terms of the formalism used to write the effective Hamiltonian (\ref{HK}) and (\ref{HKP}), 
this term is rewritten as
\begin{equation}
 V=v_{F}^2 m \sigma_{z}\,,
\end{equation}
and it corresponds to the presence of a mass term in the Hamiltonian. This type of term can be generated by covering the
surface of graphene with gases molecules \cite{ricardo} or by depositing graphene on top
of boron nitride \cite{BN0,BN1,zecarlos}. The eigenvalues of $H_{\boldsymbol{K}}+V$ are easily obtained, leading to
$
E = \pm \sqrt{v_{F}^{2} \hbar^{2} | k |^{2} + m^{2} v_{F}^{4}}$, with $\vert k\vert=\sqrt{k_x^2+k_y^2}$,
and the same for $H_{\boldsymbol{K'}}+V$.

\section{Confinement of Dirac electrons on a strip}

Our goal in this section is on deriving a mathematical framework for describing 
 the effect of confinement on
Dirac electrons. The confinement can be produced either
by etching, by the reduced dimensions of the graphene crystallites, or by
the application of gate
potentials (here the Klein tunneling poses strong limitations on the
use of such method).

\subsection{Boundary conditions and transverse momentum quantisation}

The mathematical description of the confinement requires to impose
appropriate boundary conditions to the Dirac fermions. Although,
for graphene ribbons, the two types of edges discussed above impose two
different types of boundary conditions \cite{brey}, we shall use here
the infinite mass confinement
\footnote{It is important to comment here on this particular
choice for the boundary condition. Using the $\bm k\cdot \bm p$ approach of 
DiVincenzo and Mele \cite{mele} one learns that $v_F\propto 1/m$, where $m$ is the
bare electron mass. On the other hand, one considers that graphene electrons
can not propagate in a region where the material is absent, and therefore have
zero velocity there. Due to the  proportionality $v_F\propto 1/m$, this can be achieved
taking the limit $m\rightarrow\infty$. Therefore we could think that confinement of Dirac fermions
could be achieved with a position dependent Fermi velocity $v_F(y)$, that goes to zero
at the edge of the strip. Unfortunately this program does not work.
Berry and Mondragon boundary boundary condition\cite{berry}
corresponds to a change in the nature of the spectrum, and is somewhat artificial in what
concerns graphene.
The consequences of the different boundary conditions are:
 the properties of the wave function at graphene edges do depend on the
 different choices of boundary conditions, but the 
bulk behavior of the electronic states is essentially the same for all them;
very close to the neutrality point the choice of boundary condition do again matter,
but at finite doping this is not the case anymore. The choice
of Berry and Mondragon boundary condition\cite{berry} does introduce a certain
degree of simplicity in the calculations.}
proposed by Berry and Mondragon
\cite{berry}. For large ribbons, there will be no important difference
between the two types of boundary conditions \cite{akhmerov}, except
that the infinite mass boundary condition is not able to produce edge
states \cite{PRLeduardo}, which are present in ribbons with zigzag
edges.

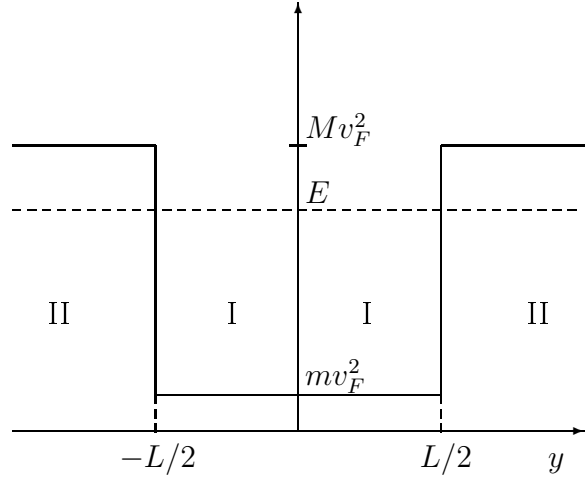
\begin{figure}[t]
\centering
\setlength{\unitlength}{0.95mm}
\begin{picture}(100,80)
\put(50,10){\vector(0,1){60}}
\put(10,10){\vector(1,0){80}}
\put(70,50){\line(1,0){20}}
\put(30,50){\line(-1,0){20}}
\put(51,51){$M v_{F}^{2}$}
\put(48.8,50){\line(1,0){2.4}}
\put(70,50){\line(0,-1){35}}
\multiput(70,15)(0,-2){3}{\line(0,-1){1.2}}
\put(30,50){\line(0,-1){35}}
\multiput(30,15)(0,-2){3}{\line(0,-1){1.2}}
\put(82,25){II}
\put(15,25){II}
\put(59,25){I}
\put(40,25){I}
\put(51,16.5){$m v_{F}^{2}$}
\put(30,15){\line(1,0){40}}
\put(51,42){$E$}
\multiput(10,41)(2.1,0){38}{\line(1,0){1.2}}
\put(85,5){$y$}
\put(67,5){$L / 2$}
\put(25,5){$- L / 2$}
\end{picture}
\caption{Scheme of the mass confinement (along $y$) with mass $m$ inside the strip and mass
$M$ outside.} \label{fig2}
\end{figure}

We shall generalise some of the results of
Refs. \cite{beenakker06,berry} by considering the case of Dirac
fermions with a finite mass. The mass profile in the transverse
direction ($y$) of the strip is represented in Fig. \ref{fig2}.
The boundary conditions the wave function has to obey, at the spatial point where the
mass changes from $m$ to $M$,
are derived considering the reflection of the wave function at the boundary. Let us first consider
the reflection at $y=L/2$. The wave function in the central region ($I$) is given by 
\begin{equation}
\psi_{I}(x,y) = \Bigg[ \left ( \begin{array}{ccc} 1 \\ f_{I}(E) e^{i \theta_{k}} \end{array} \right ) e^{i k_{y} y} + R \left ( \begin{array}{ccc} e^{i \theta_{k}} f_{I}(E)^{-1} \\  1 \end{array} \right ) e^{- i k_{y} y} \Bigg] e^{i k_{x} x},
\end{equation}
with $\theta_{k}=\arctan(k_y/k_x)$,
\begin{equation} \label{eq:fI}
f_{I}(E) = \frac {E - m v_{F}^{2}}{\sqrt{E^{2} - m^{2} v_{F}^{4}}}.
\end{equation}
 In zone $II$ ($y>L/2$), the solution has the form

\begin{equation}
\psi_{II}(x,y) =  T \left ( \begin{array}{ccc} 1 \\ f_{II}(E) \end{array} \right ) e^{i q_{y} y}  e^{i k_{x} x},
\end{equation}
with
\begin{equation}
f_{II}(E) = \frac {E - M v_{F}^{2}}{v_{F} \hbar (k_{x} - i q_{y})}\,,
\end{equation}
and
\begin{equation}
q_{y} = \pm \sqrt{ \frac{E^{2} - M^{2} v_{F}^{4}}{v_{F}^{2} \hbar^{2}} - k_{x}^2}\,,
\end{equation}
 where the energy values are given by the same expression as that for zone $I$.
As we want to take the limit $M \rightarrow \infty$, we will assume $M v_{F}^{2} > E$, which implies that
\begin{equation}
q_{y} = \pm i \sqrt{ \frac{ M^{2} v_{F}^{4} - E^{2}}{v_{F}^{2} \hbar^{2}} + k_{x}^2} = \pm i |q_{y}|,
\end{equation}
and thus
\begin{equation}
f_{II}(E) = \frac {E - M v_{F}^{2}}{v_{F} \hbar (k_{x} + |q_{y}|)} = \frac {E - M v_{F}^{2}}{v_{F} \hbar (k_{x} \pm \sqrt{ \frac{ M^{2} v_{F}^{4} - E^{2}}{v_{F}^{2} \hbar^{2}} + k_{x}^2 })},
\end{equation}
where the sign $\pm$ in front of the square root applies to the wave function that is propagating in the positive/negative 
 $y$ direction.. 
Imposing  the boundary condition associated to the Dirac equation for a reflection at $y =L/2$,
\begin{equation}
\psi_{I}(x,\frac{L}{2}) = \psi_{II}(x,\frac{L}{2})\,,
\end{equation}
one obtains 
\begin{equation}
\frac{\psi_{I_{1}}}{\psi_{I_{2}}} = \frac{\psi_{II_{1}}}{\psi_{II_{2}}} = \frac{1}{f_{II}(E)}\,.
\end{equation}
Taking the limit $M\rightarrow\infty$
the boundary conditions reduce to 
\begin{eqnarray}
\frac{\psi_{I_{1}}}{\psi_{I_{2}}} \Bigg|_{y = - L / 2} & = & + 1, \\
\frac{\psi_{I_{1}}}{\psi_{I_{2}}} \Bigg|_{y = L / 2} & = & - 1\,.
\label{bcM}
\end{eqnarray}

\begin{figure}[!hbp]
\centering
\includegraphics[width=7cm]{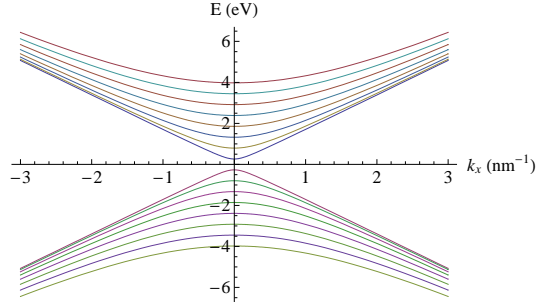}
\caption{(colour on-line) Energy levels of a ribbon 10 nm wide. The transverse modes range from $n = 0$ to $n = 8$, for both  particles and  holes.} \label{fig:RibbonLevels}
\end{figure}

Now one wants to write down the wave function of electrons propagating on the strip taking into
account the confinement due to the mass term. The most general wave function is of the sum
of two counter-propagating waves in the $y$ direction
\begin{eqnarray}
\psi(x,y) & = & \chi (y) e^{i k_{x} x}\,, 
\end{eqnarray}
where
\begin{eqnarray}
\chi (y) & = & A \left ( \begin{array}{ccc} 1 \\ f_I(E) e^{i \theta_{k}} \end{array} \right ) e^{i k_{y} y} + B \left ( \begin{array}{ccc} 1 \\ f_I(E) e^{- i \theta_{k}} \end{array} \right ) e^{- i k_{y} y}.
\end{eqnarray}
It is always possible to redefine $B$ such that

\begin{eqnarray}
\chi (y) =   A\left ( \begin{array}{ccc} 1 \\ f_I(E) e^{i \theta_{k}} \end{array} \right ) e^{i k_{y} y} + B \left ( \begin{array}{ccc} e^{ i \theta_{k}} f_I(E)^{-1} \\ 1 \end{array} \right ) e^{- i k_{y} y} \,,
\label{eq:generalSolution}
\end{eqnarray}
a procedure that proves useful later on. Imposing the boundary conditions (\ref{bcM}),
one obtains (considering the strip to be in the range $0<y<L$ for simplicity)
\begin{equation}
B = \frac{1 - f_{I}(E) e^{i \theta_{k}}}{1 - f_{I}(E)^{-1} e^{i \theta_{k}}} A\,,
\label{eq:B}
\end{equation}
for the relation between the coefficients, and
\begin{eqnarray}
e^{i k_{y} L} = - \frac{1 - f_{I}(E) e^{i \theta_{k}}}{1 - f_{I}(E)^{-1} e^{i \theta_{k}}} \frac{1 +  f_{I}(E)^{-1} e^{i \theta_{k}}}{1 + f_{I}(E) e^{i \theta_{k}}} \label{eq:expConstraint}
\label{momentum}
\end{eqnarray}
for the energy quantisation.
It is clear that the admissible values of $k_y$ are energy dependent. Considering the limit $M\rightarrow\infty$
one obtains, from Eq. (\ref{eq:B}), the simpler result $A=B$, and, from Eq. (\ref{momentum}),
the condition $e^{i 2 k_{y} L} = - 1$, 
which leads to the transverse momentum quantisation rule \cite{berry}
\begin{eqnarray}
k_{y_{n}} = \frac{\pi}{2 L} + \frac{n \pi}{L} \textrm{ , where } n = 0, \pm 1, \pm 2,\ldots \,.
\end{eqnarray}
Putting all together, the obtained results  are summarised as:

\begin{eqnarray}
\label{psi}
\psi_{n,k} (x,y) &=& \chi_{n,k} (y) e^{i k x}, \nonumber \\
\nonumber \\
\chi_{n,k} (y) &=& A \Bigg[ \left ( \begin{array}{ccc} 1 \\ z_{n,k} \end{array} \right ) e^{i q_{n} y} + \left ( \begin{array}{ccc}  z_{n,k} \\ 1 \end{array} \right ) e^{- i q_{n} y} \Bigg], \label{eq:UnnormalizedChi}
\end{eqnarray}

\noindent where we have used $k = k_{x}$, $q_{n} = k_{y_{n}}$, $s=\pm 1=\textrm{sign}[E]$ and

\begin{eqnarray}
z_{n,k} = s e^{i \theta_{k}}  = s \frac{k + i q_{n}}{\sqrt{k^{2} + q_{n}^{2}}}. \nonumber
\end{eqnarray}
If we ignore questions of convergence,
we recognise that this form for  $z_{n,k}$ does not require $k$ or $q_n$ to be
real. We are only assuming $k^{2}+q_n^{2}>0.$
The dependence of the energy $E$ on $k$ shows a number of sub-bands separated by energy gaps; this is shown
in Fig. (\ref{fig:RibbonLevels}) for a ribbon 10 nm wide. It is clear that for such a narrow ribbon one
has large energy gaps between two consecutive sub-bands.

Let us represent
Eq. (\ref{psi}) as $|\Psi_{n,k} \rangle = |\chi_{n,k} \rangle e^{i k x}$, it is then simple to show that 
$\langle \Psi_{m,k'} | \Psi_{n,k} \rangle=0$ and that the normalisation coefficient $A$ reads 
$A =1/(2 \sqrt{L})$.
Note that if on the strip we have a non-zero scalar potential $\hat{1}
V$, we will just have to substitute $E$ by $E - V$ and  replace $k$ by $\tilde{k}$, with $\tilde{k}$
given by
\begin{equation}
\tilde{k}^{2}=\frac {(E-V)^2}{v_{F}^{2} \hbar^{2} } - \frac {v_{F}^{2} m^{2} }{ \hbar^{2} }-q_n^{2}\,.
\end{equation}

\subsection{Dirac fermions in a strip with a step potential}

\begin{figure}[ht]
\centering
\setlength{\unitlength}{0.65mm}
\begin{picture}(200,95)

\put(25,10){\vector(1,0){170}}
\put(110,10){\vector(0,1){70}}
\put(110,10){\vector(-1,1){30}}

\put(5,30){\line(1,0){85}}
\multiput(90,30)(2,0){41}{\line(1,0){1.1}}
\put(172,30){\vector(1,0){1}}
\put(90,30){\line(0,1){30}}
\multiput(90,60)(0,2){19}{\line(0,1){1.1}}
\put(90,98){\vector(0,1){1}}
\put(110,40){\line(-1,1){20}}
\put(90,60){\line(1,0){75}}
\put(110,40){\line(1,0){75}}

\multiput(11,30)(0,2){32}{\line(0,1){1.1}}
\multiput(31,10)(0,2){32}{\line(0,1){1.1}}
\multiput(5,90)(2,0){80}{\line(1,0){1.1}}
\multiput(25,70)(2,0){80}{\line(1,0){1.1}}
\multiput(178,40)(0,2){17}{\line(0,1){1.1}}
\multiput(158,60)(0,2){17}{\line(0,1){1.1}}
\put(158,60){\line(0,-1){30}}
\put(178,40){\line(0,-1){30}}

\put(190,5){$x$}
\put(75,35){$y$}
\put(105,35.5){$V$}
\put(87,24.5){$L$}
\put(108.5,5){$0$}

\end{picture}
\caption{Representation of a step potential in a strip with lateral confining infinite mass.
Zone I is for $x<0$ and zone II is for $x>0$.} 
\label{fig3}
\end{figure}
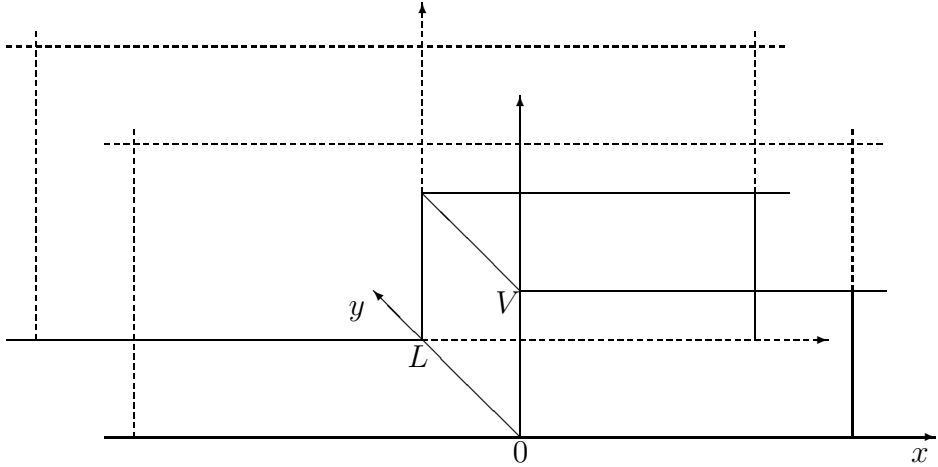

\begin{figure}[hb]
\centering
\includegraphics[width=8cm]{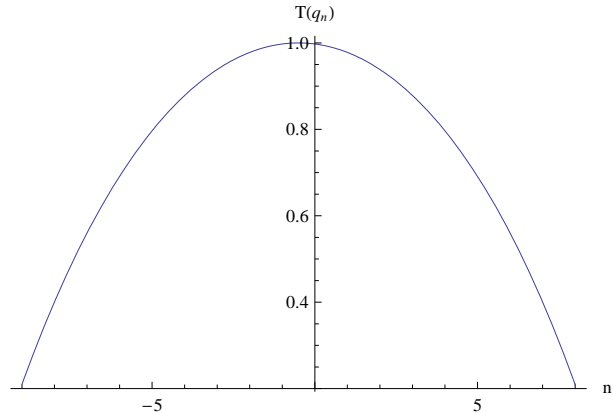}
\caption{\label{fig_step}
Transmission  coefficient, $T(q_n)=\vert t_n\vert^2$, through an energy barrier, of length $w=100$ nm and
height $V=50$ eV, as function of the transverse
quantisation quantum number $n$. The energy of the electron is taken as $E=0.1$ eV
and the width of the ribbon is $L=500$ nm.
}
\end{figure}

Let us now consider the scattering of Dirac fermions in a strip by a simple step potential, as represented
in Fig. \ref{fig3}.
In the zone $I$ one has  $V = 0$, the wave function is given by Eq. (\ref{psi});  
in zone $II$, with $V>0$, the wave function is also given by Eq. (\ref{psi}) making the
replacement $k\rightarrow \tilde k$. 
If, in general, the step rises up  at $x=X$ the boundary condition takes the form
\begin{eqnarray}
\psi_{n,k}(X,y) + r_n \psi_{n,-k}(X,y) &=& t_n \psi_{n,\tilde{k}}(X,y)\,.
\end{eqnarray}
Solving for $r$ and $t$ gives

\begin{eqnarray}
\label{r}
r_n &=& \frac{z_{n,k}^{2} - z_{n,k} z_{n,\tilde{k}}}{1 + z_{n,k} z_{n,\tilde{k}}} e^{2 i k X}\,, \\
\label{t}
t_n &=& \frac{1 + z_{n,k}^{2}}{1 + z_{n,k} z_{n,\tilde{k}}} e^{- i (\tilde{k} - k) X}\,.
\end{eqnarray}
Equations (\ref{r}) and (\ref{t}) represent the reflection and the transmission amplitudes,
respectively, for the transverse mode $n$. It is now a simple matter to compute the tunnelling
transmission for an arbitrary configuration of finite potential steps by using these
two results combined with a transfer matrix method \cite{zecarlos}. 
A particular case of this situation is the transmission through
a potential barrier, for which $\vert t_n\vert^2$ is given in Fig. \ref{fig_step}.

\subsection{Trapped eigenmodes}
\begin{figure}[ht]
\centering
\setlength{\unitlength}{0.64mm}
\begin{picture}(220,90)

\put(5,15){\vector(1,0){90}}
\put(30,5){\vector(0,1){50}}

\put(5,40){\line(1,0){85}}
\put(70,15){\line(0,1){25}}

\put(90,10){$x$}
\put(25,51){$y$}
\put(26,10){$0$}
\put(68,10){$w$}
\put(25,41){$L$}

\put(43,29){\footnotesize Zone I}
\put(41.5,24){\footnotesize ($V = 0$)}
\put(76,29){\footnotesize Zone II}
\put(75.5,24){\footnotesize ($V \neq 0$)}
\put(5.2,29){\footnotesize Zone III}
\put(5.5,24){\footnotesize ($V \neq 0$)}

\put(105,10){\vector(1,0){100}}
\put(125,5){\vector(1,1){35}}
\put(130,35){\vector(0,1){7}}

\multiput(108,10)(0,2){27}{\line(0,1){1}}
\multiput(128,30)(0,2){27}{\line(0,1){1}}
\multiput(195,10)(0,2){27}{\line(0,1){1}}
\multiput(215,30)(0,2){27}{\line(0,1){1}}

\multiput(105,59)(2,0){47}{\line(1,0){1}}
\multiput(125,79)(2,0){47}{\line(1,0){1}}

\put(150,30){\line(1,0){20}}
\multiput(125,30)(2,0){13}{\line(1,0){1}}
\multiput(170,30)(2,0){10}{\line(1,0){1}}
\put(190,30){\line(1,0){25}}

\put(130,10){\line(0,1){25}}
\put(105,35){\line(1,0){25}}
\put(150,30){\line(0,1){25}}
\put(125,55){\line(1,0){25}}
\put(130,35){\line(1,1){20}}

\put(170,10){\line(0,1){25}}
\put(170,35){\line(1,0){27}}
\multiput(190,35)(0,2){10}{\line(0,1){1}}
\put(190,30){\line(0,1){5}}
\put(190,55){\line(1,0){27}}
\put(170,35){\line(1,1){20}}
\put(170,10){\line(1,1){20}}

\put(198,5){$x$}
\put(152,37){$y$}
\put(129.5,5){$0$}
\put(168,5){$w$}

\put(128,22){\line(1,0){4}}
\put(122.5,20.5){$E$}

\end{picture}
\caption{Scheme of the confinement (along y) in a strip where a scalar potential well,
of width $w$, was created. On the left one has an upper view, and on the right one has a side view.} 
\label{fig4}
\end{figure}
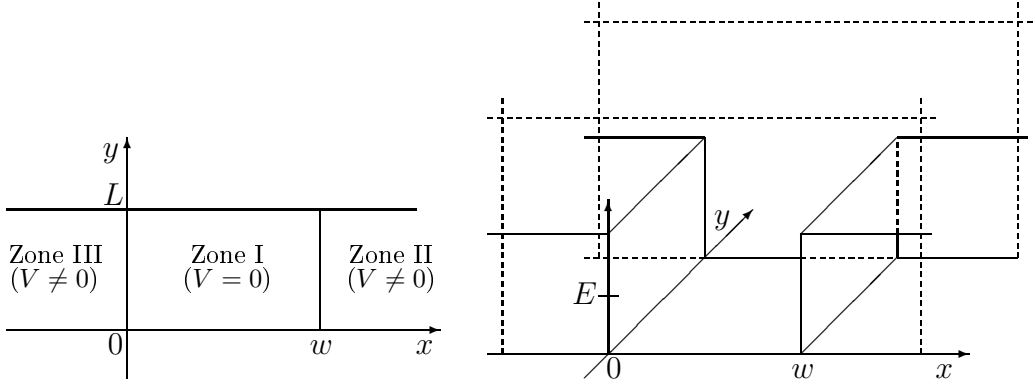

In this section we will show that it is possible to trap massless Dirac
electrons in a ribbon of finite width $L$ by creating a $p-n-p$
junction \cite{cheianov}. 
The effect exploits the fact that the spectrum of a finite
ribbon exhibits energy gaps.  A similar study was done in
Ref. \cite{TrappedEigenmodes} for the bulk case. 
In this case,  such
a confinement is possible for certain incident angles
 of the eigenmodes on the potential walls \cite{TrappedEigenmodes}.
The potential profile considered
is shown in Fig. \ref{fig4}.  We show that the trapping of  the eigenmodes requires
evanescent modes in the $x$ direction. This can be
accomplished using a scalar potential.

In order to solve this problem let as again consider the case of a potential step as in Fig. \ref{fig4}.
In region $I$ ($x<w$) the wave function has the form (\ref{psi}) and in region $II$ ($x>w$) the form
would be the same with $k$ replaced by 
\begin{equation}
\tilde{k} = \sqrt{\Bigg( \frac{E - V}{v_{F} \hbar} \Bigg)^{2} - q_{n}^{2}}.
\end{equation}
One now makes the observation that if $q_{n}$ obeys the condition
\begin{eqnarray} 
\frac{(E - V)^{2}}{v_{F}^{2} \hbar^{2}} < q_{n}^{2} < \frac{E^{2}}{v_{F}^{2} \hbar^{2}}, 
\label{eq:Condition}
\end{eqnarray}
one has a propagating wave in region $I$ and an evanescent wave in region $II$. Within the validity of 
Eq. (\ref{eq:Condition}) it is more transparent to write the wave function in region $II$ as

\begin{eqnarray}
\psi_{II_{n,\tilde{k}}}(x,y) = \frac{1}{2 \sqrt{L}} \Bigg[ \left ( \begin{array}{ccc} 1 \\ i s' \frac{\alpha + q_{n}}{\sqrt{q_{n}^{2} - \alpha^{2}}} \end{array} \right ) e^{i q_{n} y} + \left ( \begin{array}{ccc} i s' \frac{\alpha + q_{n}}{\sqrt{q_{n}^{2} - \alpha^{2}}} \\ 1 \end{array} \right ) e^{- i q_{n} y} \Bigg] e^{- \alpha x}, 
\end{eqnarray}
with 
\begin{equation}
\tilde{k} = i \alpha = i \sqrt{q_{n}^{2} - \Bigg( \frac{E - V}{v_{F} \hbar} \Bigg)^{2}},
\end{equation}
where $s' = {\rm sign}[E - V]$. Of course, the same type of analysis
holds if one had considered the step at $x=0$, starting at the
interface between regions $III$ and
$I$. The trapping ``mechanism'' uses this  fact. The wave
function in region $I$ of Fig. \ref{fig4} is taken as a sum of two
counter-propagating waves along the $x$ direction, whereas in regions
$II$ and $III$ only evanescent waves exists. Imposing the boundary
conditions at $x = 0$ and $x = w$ ($w$ the width of the well) one
obtains after a lengthy calculation the condition of the energy of
the trapped eigenmodes
\begin{eqnarray}
\sin (kw) { F}(E,V,q_{n}) + \cos (kw) { G}(E,V,q_{n}) = 0 \,,
\label{trapmodes}
\end{eqnarray}
with 

\begin{eqnarray}
 {F}(E,V,q_{n})=&=& \frac{i}{16} 
\Big[ - 4 (z_{n,k}^{*} + z_{n,k})^{2} - ( z_{n,k} - z_{n,k}^{*} ) ( z_{n,\alpha} + z_{n,-\alpha}) - \nonumber \\
&-& ( z_{n,k}^{3} - (z_{n,k}^{*})^{3} ) ( z_{n,\alpha} + z_{n,-\alpha} ) \Big]\,,
\end{eqnarray}
and
\begin{eqnarray}
{G}(E,V,q_{n}) = 
\frac{3}{16} ( z_{n,k} + z_{n,k}^{*} + z_{n,k}^{3} + (z_{n,k}^{*})^{3} ) ( z_{n,\alpha} - z_{n,-\alpha})\,.
\end{eqnarray}
Both ${F}(E,V,q_{n})$ and ${G}(E,V,q_{n})$ are pure
imaginary numbers, as long as condition (\ref{eq:Condition}) holds
true. In order to give a flavor of the numerical solution of
Eq. (\ref{trapmodes}) we present its numerical solution in Table
\ref{tab1}. We have
chosen the strategy of fixing the energy and looking for the values
of $w$ that satisfy Eq. (\ref{trapmodes}), for different values of
$q_n$.

\begin{table}
\caption{\label{tab1}Values of $w$ (in nm) for a given momentum $q_n$ (in 1/nm).
The parameters are $s = 1$, $s' = -1$, $v_{F} = 10^6$ m/s, $E = 0.1$ eV, $V = 0.15$ eV, 
and $L = 500$ nm.}
\begin{indented}
\item[]\begin{tabular}{@{}llllllll}
\br
$n$& $w$ & $n$& $w$ &$n$& $w$ &$n$& $w$  \\
\mr
5  &  $\sim$59& 6  &  $\sim$65 & 7  &  $\sim$78 & 8  &  $\sim$113\\
\br
\end{tabular}
\end{indented}
\end{table}

\section{Inducing mode mixing by scattering at a wall}

In this section we want to discuss the scattering of Dirac electrons
when they propagate along a semi-infinite narrow channel and scatterer back at the
wall located at the end of the channel. This problem
is intimately related to the possibility of finding a solution for the eigenmodes
and eigenstates of trapped Dirac electrons in a square box. Our analysis hints at
the reason why this solution has not been found yet.

\subsection{Definition of the problem}

Let us now consider massless Dirac fermions confined in a semi-infinite strip:
$x<0$ and $0<y<L$. We will look for scattering states produced by the
scattering at the wall due to an incoming wave
from $x\to-\infty$. As before, the fermions are confined by an infinite mass term outside the strip. 

The scattering state is a sum of an incoming wave, with energy $E$, longitudinal momentum
$k$, and transverse momentum $q_n$, with
a superposition of all the
possible outgoing channels with reflection amplitude $r_{n,m}$, and it can be written as
\begin{eqnarray}
\Psi_{n,k}(x,y) & = &\left(\begin{array}{c}
\Psi_1(x,y)\\
\Psi_2(x,y)\end{array}\right) =\left[\left(\begin{array}{c}
1\\
z_{n,k}\end{array}\right)e^{iq_{n}y}+\left(\begin{array}{c}
z_{n,k}\\
1\end{array}\right)e^{-iq_{n}y}\right]e^{ikx}\nonumber \\
 & + & \sum_{m=0}^{\infty}r_{n,m}\left[\left(\begin{array}{c}
1\\
z_{m,-k_m}\end{array}\right)e^{iq_{m}y}+\left(\begin{array}{c}
z_{m,-k_m}\\
1\end{array}\right)e^{-iq_{m}y}\right]e^{-ik_mx}.\nonumber \\
\label{eq:scattering_wf}
\end{eqnarray}
Since we are considering  elastic scattering ($\Psi$ is an eigenstate), we must
have \[
E^{2}=k_m^{2}+q_{m}^{2}=k^{2}+q_{n}^{2},\]
i.e., 
\begin{equation}
k_m^{2}=k^{2}+q_{n}^{2}-q_{m}^{2}.
\label{eq:reflect_k}
\end{equation}
 We must distinguish two situations: 
\begin{eqnarray}
q_{m}^{2}<E^{2} & \Rightarrow & k_m=\sqrt{E^{2}-q_{m}^{2}}\,,
\label{eq:propagating}\\
q_{m}^{2}>E^{2} & \Rightarrow & k_m=i\sqrt{q_{m}^{2}-E^{2}}\,.
\label{eq:evanescent}
\end{eqnarray}
The second case corresponds to evanescent modes. We must choose this
solution if we want the wavefunction to be convergent for $x\to-\infty$;
for real $k_m$ the choice of sign reflects the fact that we have
only one incoming mode.  We now simplify the notation, using the fact
that $k$ and $n$ are fixed, and define 
\begin{eqnarray}
z_{n,k} & = & \frac{k+iq_{n}}{s\sqrt{k^{2}+q_{n}^{2}}}=z_{n},\label{eq:zn}\\
z_{m,-k_m} & = & \frac{-k_m+iq_{m}}{s\sqrt{k^{2}+q_{m}^{2}}}=\tilde{z}_{m}.
\label{eq:ztildem}
\end{eqnarray}
The parameter $z_{n}$ is just a phase, $\left|z_{n}\right|^{2}=1$;
$\tilde{z}_{m}$ is also a phase for propagating modes, but for evanescent
modes $\left|\tilde{z}_{m}\right|^{2}\neq1$. For $q_{m}^{2}\gg E^{2}$,
we obtain
\[
\tilde{z}_{m}\approx\frac{-iq_{m}(1-E^{2}/2q_{m}^{2})+iq_{m}}{E}
\approx i\frac{E}{2q_{m}^{2}}\to0\,.\]
\subsection{Calculation of the reflection coefficients}

The boundary condition at the end of the semi-infinite strip $x=0$
and $0<y<L$, is (see Sec. \ref{secdot} for details)
\begin{equation}
 \Psi_1(0,y)+i\Psi_2(0,y)=0\,.
\label{BCy}
\end{equation}
Note that the boundary condition at an infinite-mass vertical-wall is different from that
of the horizontal case discussed before. Applying the boundary condition (\ref{BCy}) 
to $\Psi_{n,k}(x,y)$ and after rather lengthy algebra (where the replacement
$-(m+1)=m'$ is made at some stage and $m'$ is redefined as $m$ afterwards)
 one arrives at the condition
\begin{equation}
\sum_{m=-\infty}^{\infty}I(n,m)e^{iq_{m}y}=0\,,
\label{eq:bceqs_i}
\end{equation}
with $I(n,m)$ given by
\begin{eqnarray}
I(n,m) & \equiv & \left[\left(1+iz_{m}\right)\delta_{n,m}+r_{n,m}(1+i\tilde{z}_{m})\right]
\theta(m+1/2)\nonumber \\
 & + & \left[\left(z_{-m-1}+i\right)\delta_{n,-m-1}+
r_{n,-m-1}(\tilde{z}_{-m-1}+i)\right]\theta(-1/2-m)\,.
\label{eq:I_def}
\end{eqnarray}
The crucial step in the derivation is the observation that 
the set of function's $\{\phi_{m}=e^{iq_{m}y},m=0,\pm1,\pm2,\dots\}$
is \emph{overcomplete}. In fact, the set of states with $m$ even
(or with $m$ odd) is, by itself, a complete orthogonal set for functions
defined in the interval $0<y<L$. If follows that we obtain an equivalent
set of conditions to Eqs. (\ref{eq:bceqs_i}) by taking inner products
of Eq. (\ref{eq:bceqs_i}) with $\phi_{m}$ with $m$ even (or $m$ odd).
We recall the inner products (choosing $p$ even) to be
\begin{equation}
\frac{1}{L}\int_{0}^{L}dye^{-i\left(q_{p}-q_{m}\right)y}=
\left\{
\begin{array}{cc}
1, 			& p=m\\
0, 			& m\,\textrm{is even},\, m\neq p\\
\frac{2}{i(p-m)\pi},  	& m\,\textrm{is odd}
\end{array}
\right.\,.
\label{eq:inner_prodcs}
\end{equation}
The fact that the integral (\ref{eq:inner_prodcs}) is not a Kronecker symbol
shows that, in this case, the basis is overcomplete.
Using Eq. (\ref{eq:inner_prodcs}), we obtain
\begin{equation}
I(n,p)+\sum_{m_{\rm odd}}\frac{2}{i(p-m)\pi}I(n,m)=0\qquad p=0,\pm2\pm4,\dots
\label{eq:central}
\end{equation}
which is the central result of this section. 
This is the set of equations needed to calculate the  $r_{n,m}$ coefficients.
Naturally the convergence of the sum in (\ref{eq:central}) critically depends 
on the behaviour of $r_{n,m}$ with $m$.

We could also have formulated the
scattering problem a bit more generally, considering, for example, the case of a wave that approaches
a wall at $x=-D$ coming from $x=+\infty$. Naturally the modifications relatively to the
solution found before cannot be much, given the symmetry of the problem.
The first thing to note is that the boundary condition is slightly changed,
being given by (see Sec. \ref{secdot} for details)
\begin{equation}
 \Psi_1(-D,y)-i\Psi_2(-D,y)=0\,.
\label{BCyII}
\end{equation}
Working out the problem along the same lines as before, one learns that
the final result can be obtained from the previous solution upon the
replacements
\begin{eqnarray}
\label{t1}
r_{n,m}&\rightarrow r_{n,m}e^{iD(k+k_m)}\,,\\
\label{t2}
z_m&\rightarrow  -1/z_m\,,\\
\label{t3}
\tilde z_m&\rightarrow  -1/\tilde z_m\,, 
\end{eqnarray}
where the transformation (\ref{t1}) is obtained using the generator of translations,
$\hat T(x_0)=e^{ix_0\hat p/\hbar}$ (with $\hat p$ the momentum operator),
and follows from the new position of the wall. The
transformations (\ref{t2}) and (\ref{t3}) follow from the difference in the
boundary condition between a right and a left vertical wall. One should note
that the transformation (\ref{t1}) is not a phase for evanescent waves.
Using transformations (\ref{t1})-(\ref{t3}) in Eq. (\ref{eq:I_def}) one obtains

\begin{eqnarray}
&I(n,m) \rightarrow  \left[\left(1+i/z_{m}\right)\delta_{n,m}+
e^{iD(k+k_m)}r_{n,m}(1+i/\tilde{z}_{m})\right]
\theta(m+1/2)\nonumber \\
 & -  \left[\left(1/z_{-m-1}+i\right)\delta_{n,-m-1}+
e^{iD(k+k_{-m-1})}r_{n,-m-1}(1/\tilde{z}_{-m-1}+i)\right]
\theta(-1/2-m)\,.
\label{eq:I_defb}
\end{eqnarray}

Naturally, the exact solution of the scattering problem rests upon the possibility of 
solving exactly the set of linear equations (\ref{eq:central}). This task seems out of
reach at the moment. The second approach is to solve numerically this set of equations.
This leads to the conclusion that the summation over $m$ has to be truncated at some value.
To be concrete, let us consider the particular case of an incoming mode with transverse quantum number
$n=0$, such that the value of the incoming longitudinal momentum $k$ 
originates $N_p$ propagating modes above the mode $n=0$
($0<N_p<N$). Then the numerical solution of the problem has to satisfy the
conservation of the probability density current (it must be one in this case) and
the retained coefficients after the truncation have to converge upon increasing $N$.
As we show below, both these two conditions are satisfied for  small $N$.
For the numerical solution we choose
$n=0$, $q_0=\pi/2$, such that the value of $k=2\pi$ leads to $n=0$
and  $n=1$ as the only two propagating modes. We use a set of units such that
$L=1$ and $\hbar v_F=1$. Further we take $E>0$ which implies that $s=1$,
since the scattering being elastic can not excite hole states, which have
$E<0$. The equations to be solved numerically are given in \ref{ap0}.

\begin{figure}[ht]
\centering
\includegraphics*[width=8cm]{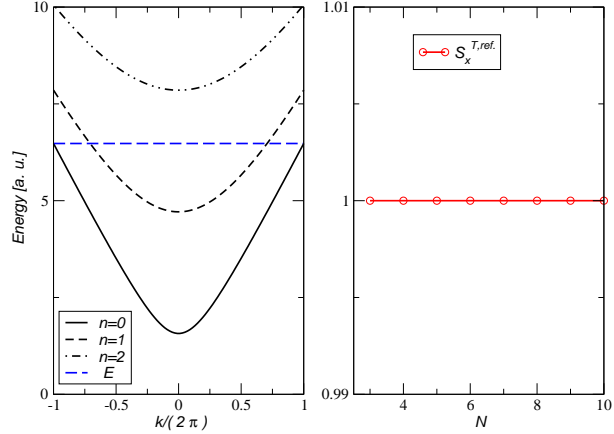}
\caption{\label{fig:scat} (colour on-line)
Left panel: Energy levels for $n=0,1,2$, and
$k=2\pi$. This leads to $N_p=1$. Right panel:
sum rule (\ref{sumrule}) for $N_p=1$ as function of $N=3,4,\ldots, 10$.  
We have depicted only odd values of $N$,
such that the total number of odd and even terms is the same ($n=0$ is considered
even), but our results are independent of this choice.}
\end{figure}

The probability density current transported by the mode $n$ is defined as
\begin{equation}
 \bm S_n(x,y)=v_F\psi^\dag_n(x,y)\bm\sigma\psi_n(x,y)\,.
\label{S}
\end{equation}
Since the motion is transversely confined, what is needed is the probability density current 
along the $x$-direction, which reads
\begin{equation}
 S_n(x)=v_F \int_0^L dy\,\psi^\dag_n(x,y)\sigma_x\psi_n(x,y)\,.
\label{Sx}
\end{equation}
Using definition (\ref{Sx}), the total reflected flux density, $S^{T,refl}_x$, obeys the sum rule
 \begin{equation}
 S^{T,refl}_x=\vert r_{0,0}\vert^2
+\sum_{m=1}^{N_p}\frac{\cos\beta_m}{\cos\beta_0}\vert r_{0,m}\vert^2=1
\,,
\label{sumrule}
\end{equation}
with 
\begin{equation}
 \beta_n = \arctan \frac {q_n}{k_n}\,,
\end{equation}
with both ${q_n}$ and ${k_n}$ real.
In agreement with our expectations, the sum-rule is better fulfilled the larger
$N$ is, although modest values of $N$ do a good job as well. 
In fact, in the right panel of Fig. \ref{fig:scat}
one can see that the sum rule is fulfilled even considering only one
evanescent mode ($N=3$).

\begin{figure}[ht]
\centering
 \includegraphics*[width=8cm]{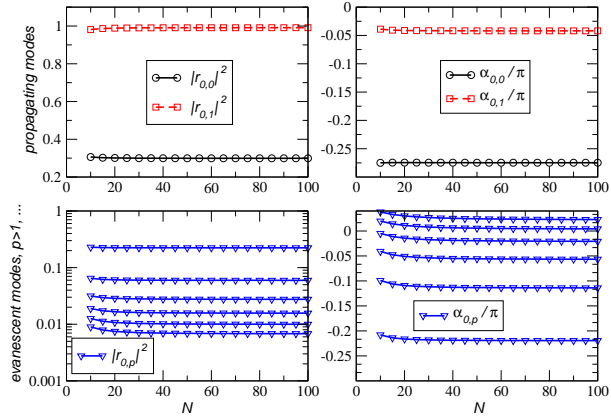}
 \caption{\label{fig:randp} (colour on-line)
Top panels: evolution of 
 $\vert r_{0,0}\vert^2$ and $\vert r_{0,1}\vert^2$ as function of $N$ (left);
 evolution of the phases $\alpha_{0,0}$ and $\alpha_{0,1}$ as function of $N$
 (right). Bottom panels: the same as before but now for $\vert r_{0,p}\vert^2$
 (left) and for $\alpha_{0,p}$ (right), considering $p=2,\ldots, 7$.
 }
 \end{figure}

In Fig. \ref{fig:randp} we study, for the particular mode occupation defined in the
left panel of Fig. \ref{fig:scat},
the evolution of the coefficients $r_{0,n}$ as function of $N$.
We write each coefficient  $r_{0,p}$ as $r_{0,p}=\vert r_{0,p} \vert e^{i\alpha_{0,p}}$.
In Fig. \ref{fig:randp}, we plot the square of the modulus of $r_{0,p}$ (left panels)
and the corresponding phase $\alpha_{0,p}$ (right panels), separating the
cases for which $p=0,1$ (propagating modes), which are represented in the two top
panels,
from those where $p\ge 2,\ldots 7$ (evanescent modes), which we represent in the two bottom panels.
A beautiful result emerges from this study. The wall introduces mode mixing and
generates evanescent waves, whose contribution to the total wave function
diminuishes upon increasing $p$. This result is quite different from that for
Schr\"odinger electrons, where no mode mixing takes place. The fundamental reason
is due to the fact that for Schr\"odinger electrons the transverse wave-function
is the same for the incoming and outgoing waves. For Dirac electrons, on the contrary,
the spinor of the incoming and outgoing waves change due to its dependence on either 
the incoming or outgoing momentum.

Had we tried to force the solution of the problem using only one incoming and
one outgoing propagating modes, with the same $k$ value, and we would have obtained the
trivial solution $k=0$. This statement is easily proved as follows:
we make the assumption that the total wave function should be a sum
of two terms of the form

\begin{eqnarray}
\Psi_{n,k}(x,y)  = \left(\begin{array}{c}
\Psi_1(x,y)\\
\Psi_2(x,y)\end{array}\right) =&
\left[\left(\begin{array}{c}
1\\
z_{n,k}\end{array}\right)e^{iq_ny}+\left(\begin{array}{c}
z_{n,k}\\
1\end{array}\right)e^{-iq_ny}\right]e^{ikx}+\nonumber \\
 &  \left[\left(\begin{array}{c}
1\\
z_{n,-k}\end{array}\right)e^{iq_ny}+\left(\begin{array}{c}
z_{n,-k}\\
1\end{array}\right)e^{-iq_ny}\right]e^{-i(kx-2\delta)}\,,\nonumber \\
\label{eq:scattering_specialcase}
\end{eqnarray}
where the phase-shift $\delta$ was introduced. Let us now impose the boundary
condition (\ref{BCy}) on the wave function (\ref{eq:scattering_specialcase}).
Working out the calculation, one obtains two conditions that must be fulfilled
simultaneously
\begin{eqnarray}
 \cos\delta \pm \sin(\beta_{n,k} -\delta)=0\,,
\label{condition}
\end{eqnarray}
with $\beta_{n,k}$  defined from $z_{n,k}=e^{i\beta_{n,k}}$. It is
clear that the two conditions in Eq. (\ref{condition}) cannot be satisfied, in general,
 at the same time, which precludes the proposal of Eq. (\ref{eq:scattering_specialcase})
as a solution to the problem. In fact, the two conditions given by 
Eq. (\ref{condition}) are equivalent to
\begin{eqnarray}
\delta = \frac{\pi}{2} +\ell\pi \wedge \delta = \beta_{n,k} + \ell\pi\,,
\hspace{1cm} \ell=0,1,2,\ldots\,,   
\end{eqnarray}
which can only be true if $2\beta_{n,k}=\pi$, a situation that occurs only if
\begin{equation}
  \arctan\frac{q_n}{k}=\frac{\pi}2\,,
\end{equation}
which finally is true only in the trivial case  $k=0$. This means that the wave function has no $x$ dependence
 and that the electronic density is $\rho(x,y)=\Psi^\dag_{n,k}(x,y)\Psi_{n,k}(x,y)=1/L$, constant everywhere
(for finite $m$ the density does show oscillations inside the box \cite{fiolhais}).
The exact solution of the square billiard with infinite-mass confinement is therefore 
a quite elusive problem \cite{berry}.

\section{Confinement of Dirac fermions in quantum dots}
\label{secdot}

\begin{figure}[ht]
\centering
\includegraphics[width=8cm]{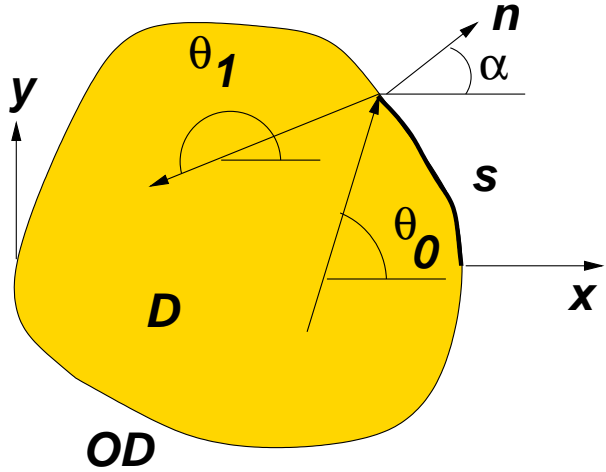}
\caption{\label{fig5}(colour on-line)
Domain $D$ with the boundary parametrised by $s$. Figure adapted from Ref. \cite{berry}.
The incident and reflected wave at $s$ are both shown.}
\end{figure}

Let us now consider the confinement of Dirac fermions in quantum
dots. Naively one would expect that the rectangular dot would have a
simple solution (as it has in the Schr\"odinger case), since the wave
function of the confined Dirac electrons (by an infinite mass term)
in a strip can be written in terms of elementary trigonometric
functions. In fact this is not the case. The only  known case so far of
an integrable Dirac dot (billiard), subjected to the infinite mass confinement, is the
circular one.  In order to solve this problem one needs the boundary
condition obeyed by the wave function at the dot boundary.  This was
worked out by Berry and Mondragon \cite{berry} and the geometry they
used is represented in Fig. \ref{fig5}. They considered a quantum dot
represented by a domain $D$ of arbitrary shape, separated by an
outside region that we denominate $OD$. The boundary of the domain $D$ is
parametrised by an length arc $s(\alpha)$, where the vector normal to
the surface of the dot at $s$ is given by
\begin{eqnarray}
{\bm n} (s) = \cos \alpha(s) \ \vec{e}_{x} \ + \ \sin \alpha(s) \ \vec{e}_{y} \label{eq:UnitNormal}
\end{eqnarray}
Imposing the condition of zero flux perpendicular to the wall of the dot one obtains
\begin{eqnarray}
\frac{\psi_{2}}{\psi_{1}} = i B e^{i \alpha (s)}\,.
\label{bcg}
\end{eqnarray} 
The constant $B$ is determined working out the study of a reflecting wave at the boundary of the dot, when the mass in the region
$OD$ obeys the condition $M\rightarrow\infty$. The final result is $B=1$ \cite{berry}, and  the
detailed calculation can be found in
\ref{ap1}. 

\subsection{The circular dot with zero magnetic field}

Let us first write the free solutions of the Dirac equation in polar coordinates
$r$ and $\f$. In this coordinates the Dirac Hamiltonian and the wave function read
\cite{hentschelprb07,recherprb07,wunschprb08,cserti,apalkov}

\begin{eqnarray}
H &=& - i \hbar v_{F} \left ( \begin{array}{cc} 0 & e^{- i \f} (\partial_{r} - \frac{i}{r} \partial_{\f}) \\ e^{i \f} (\partial_{r} + \frac{i}{r} \partial_{\f}) & 0 \end{array} \right ), 
\label{hcirc}
\end{eqnarray}
and
\begin{eqnarray}
\Psi_{k,m} (r, \f) = \left ( \begin{array}{c} J_{m} (r k) e^{i m \f} \\ s i J_{m + 1} (r k) e^{i (m + 1) \f} \end{array} \right )\,,
\label{psicirc}
\end{eqnarray} 
respectively, where $J_m(x)$ is the Bessel function of integer order $m$. 
A detailed derivation of these results is given in  \ref{ap2}.
In order to obtain the eigenvalues of the electrons in the dot one has to apply the
boundary condition (\ref{bcg}). Note that because one has a circular dot $\alpha (s)=\varphi$.
This latter property makes it possible to satisfy the boundary condition (\ref{bcg}) with the
wave function (\ref{psicirc}) alone. In fact, for a dot of radius $R$, one has
\begin{eqnarray}
s i J_{m + 1} (R k) e^{i (m + 1) \f} = J_{m} (R k) e^{i m \f}i  e^{i \f}
\Leftrightarrow
s J_{m + 1} (R k) = J_{m} (R k) \,,
\label{edot}
\end{eqnarray}
whose numerical solution gives the value of $kR$ for a given $s$ and
$m$, and from this the energy levels are computed using $E_{s,m,j} =
s\hbar v_{F} k_{s,m,j}$. The eigenvalues $E_{s,m,j}$ are defined by
three quantum numbers: $s$, $m$, and $j$, where $j$ represents the
ascending order of the values of $kR$ that satisfy (\ref{edot}), for a
given $s$ and $m$. In the next section we give numerical results
for the energy eigenvalues.

\subsection{The circular dot in a finite magnetic field}

Let us now see how we can adapt our formalism to address the calculation of the energy eigenvalues
of a circular dot in a magnetic field, that is we want to study the formation of Landau levels
in reduced geometries (amusing enough, the first calculation of Landau levels using the
Dirac equation is as old as quantum mechanics itself \cite{Rabi28}, a result that 
was forgotten by the graphene community). 
Experimentally this situation has been realised in Ref. \cite{ethdots}. The cyclotron motion of bulk graphene was discussed in Ref. \cite{Schliemann}.

The Hamiltonian (\ref{hcirc}) was written for a single Dirac cone. As is shown in 
\ref{ap2}, the Hamiltonian for the two Dirac cones can be written using an additional quantum number $\kappa=\pm$,
associated with the valley index, reading

\begin{equation}
H_\kappa=-\hbar v_{F}  \left ( \begin{array}{cc}
0&i\partial_x+\kappa\partial_y\\
i\partial_x-\kappa\partial_y&0
\end{array}\right).
\label{kinetic}
\end{equation}

In this section we do not use the infinite mass boundary condition,
but introduce the zigzag type of boundary condition. This will allow
us to consider the presence edge states \cite{WFA+99}. As we will show
in the next section, these states are always present in graphene quantum
dots.  Recalling Fig. \ref{fig1}, one sees that at the zigzag edge
only one type of carbon atom (either $A$ or $B$) is present. The
boundary condition at a zigzag edge with, say, only $B$ atoms present,
requires that the amplitude of the wave function at the $A$ atoms to
be zero, we therefore have the condition
\begin{equation}
\label{zigzag}
\psi_1(R,\f)=0\,.
\end{equation}
In the following we will choose the boundary condition
defined by Eq. (\ref{zigzag}), for which the wave vector is quantised as $k=z_{
mj}/R$, where $z_{mj}$ denotes the $j$-th root of the $m$-th Bessel
function, $J_m(z_{mj})=0$. 

A magnetic field $\vec B=B \vec e_z$, perpendicular to the graphene sheet, gives rise to a vector potential, which in polar coordinates  reads $\vec A=A_\f\vec e_\f$, and, using Gauss' theorem, one obtains $2\pi rA_\f=\pi r^2B$.  Making the traditional minimal
coupling of the charged electrons to the vector potential, the Hamiltonian has the form
\begin{equation}
H_\kappa=-i\hbar v_{F} 
\left ( \begin{array}{cc}
0&e^{-i\kappa\varphi}(\partial_r-\kappa\frac{i}{r}\partial_\varphi+\kappa\frac{\pi Br}{\Phi_0})\\
e^{i\kappa\varphi}(\partial_r+\kappa\frac{i}{r}\partial_\varphi-\kappa\frac{\pi Br}{\Phi_0})&0
 \end{array}\right)\,,
\label{HwithB}
\end{equation} 
where $\Phi_0=h/e\simeq 4136$ T$\cdot$nm$^2$ denotes the elementary flux quantum  and $-e$ is the electron charge. We now make
the observation that the trial function
\begin{equation}
\Psi_{m,\kappa}(r,\f)
= \left( \begin{array}{c}  \psi_{m,\kappa}^1(r) e^{i 
    m\f} \\ \psi_{m,\kappa}^2(r)e^{i(m +\kappa)\f} \end{array} \right)\,,
\label{AnsatzWavefunction}
\end{equation}
renders the eigenvalue problem a one-dimensional one, with the radial Hamiltonian given by
\begin{equation}
H_\kappa=-i\hbar v_F 
\left( \begin{array}{cc} 
0&\partial_r+\frac{\kappa m+1}{r}+\kappa\frac{\pi Br}{\Phi_0}\\
\partial_r-\frac{ \kappa m}{r}-\kappa\frac{\pi Br}{\Phi_0}&0
\end{array} \right)\,.
\label{HwithBRadial}
\end{equation}
Let us make the  substitution $\psi^i=\tilde\psi^i/\sqrt{r}$ ($i=1,2$)
in the eigenvalue equation defined by the Hamiltonian (\ref{HwithBRadial}), with the radial spinor wave function having the
form $\psi=(\psi^1,\psi^2)$. This substitution was considered before in the exact solution of the
Coulomb problem in the 2+1 dimensional Dirac equation \cite{Dong03} and also in \cite{wunschprb08}.
This procedure leads to a more symmetric eigenproblem of the form
\begin{eqnarray}
\label{EG1}
&-&i\hbar v_F \left[\partial_r\tilde\psi_{m,\kappa}^2(r)+\left(\frac{\kappa m+1/2}{r}+\kappa\frac{\pi Br}{\Phi_0}\right)\tilde\psi_{m,\kappa}^2(r)\right]
=E\tilde\psi_{m,\kappa}^1(r)\,,\\
\label{EG2}
&-&i\hbar v_F \left[\partial_r \tilde\psi_{m,\kappa}^1(r)-\left(\frac{\kappa m+1/2}{r}+\kappa\frac{\pi Br}{\Phi_0}\right)\tilde\psi_{m,\kappa}^1(r)\right]
=E\tilde\psi_{m,\kappa}^2(r)\,.
\end{eqnarray}
We want to solve the eigenproblem defined by (\ref{EG1}) and (\ref{EG2}) by diagonalising an Hermitian matrix.
\begin{figure}[ht]
\begin{center}
\includegraphics*[angle=0,width=12cm]{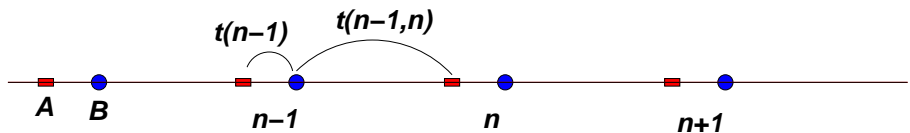}
\caption{(colour on-line) Representation of a dimerised chain of atoms $A$ and $B$.
Within the unit cell $n$ the hopping is $t(n)$ and between the unit cells $n$ and
$n+1$ the hopping is $t(n,n+1)$. They can be functions of the unit cell position
$n$.
 \label{fig6}}
\end{center}
\end{figure}
To this end let us look at the problem of a
dimerised one-dimensional tight-binding model, such as that
represented in Fig. \ref{fig6}.  The relevance of this interlude will be apparent in a moment. 
The Hamiltonian for the depicted
system is
\begin{equation}
H=\sum_n [t(n)\vert nA\rangle\langle nB\vert + t(n,n+1)\vert nB\rangle\langle n+1A\vert + {\rm H.\, c.}]\,,
\end{equation}
and the wave function is written as
\begin{equation}
 \vert \Psi\rangle =\sum_n (a_n\vert nA\rangle + b_n\vert nB\rangle)\,.
\end{equation}
The eigenvalue equation $H\vert \Psi\rangle=E\vert \Psi\rangle$ can be reduced to the solution of the
linear homogeneous system
\begin{eqnarray}
b_nt(n)+b_{n-1}t(n-1,n)=a_nE\,, \\
a_nt(n)+a_{n+1}t(n,n+1)=b_nE\,.
\end{eqnarray}
Introducing the simplifying notation $t(n,n+1)=t'(n)$, the above eigensystem reads
\begin{eqnarray}
\label{tb1}
b_nt(n)+b_{n-1}t'(n-1)&=&a_nE\,, \\
\label{tb2}
a_nt(n)+a_{n+1}t'(n)&=&b_nE\,.
\end{eqnarray}
We note that Eqs. (\ref{tb1}) and (\ref{tb2})
pose well defined numerical problem for  well behaved functions $t(n)$ and $t'(n)$.
Let us now see what kind of continuous model follows from this lattice problem.
Notice that since the model under consideration has a valence and a conduction bands, in 
case we have one electron per site the relevant energies are around zero. In this
case, the amplitudes $a_n$ and $b_n$ oscillate between positive and negative values
within a lattice unit cell. In order to construct a well defined continuous model
we need to subtract this oscillatory behaviour making the replacement
$a_n=i(-)^n\tilde a_n$ and $b_n=(-)^n\tilde b_n$. This produces the set of equations
\begin{eqnarray}
\label{tb1b}
-i[\tilde b_nt(n)-\tilde b_{n-1}t'(n-1)]&=&\tilde a_nE\,, \\
\label{tb2b}
-i[-\tilde a_nt(n)+\tilde a_{n+1}t'(n)]&=&\tilde b_nE\,.
\end{eqnarray}
Defining now $T(n)=[t(n)+t'(n)]/2$ and $\Delta(n)=[t(n)-t'(n)]/2$
and recalling that the first order derivatives can be approximated by
\begin{eqnarray}
 \partial_r\tilde a\rightarrow [\tilde a(r_{n+1})-\tilde a(r_n)]/\Delta r\,,\\
\partial_r\tilde b\rightarrow [\tilde b(r_{n})-\tilde b(r_{n-1})]/\Delta r\,.
\end{eqnarray}
and that $\Delta r=R/N_l$,  $r_n=Rn/N_l$
a
discretised position vector, with $R$ the length of the chain (which will correspond latter to the radius of the dot), 
$n=1,\ldots N_l$, and $N_l$ the number of points  in which the length $R$ was discretised, we obtain
\begin{eqnarray}
\label{tb1c}
&-&i[T(r)\Delta r\partial_r \tilde b+2\Delta(r)\tilde b]=\tilde aE\,, \\
\label{tb2c}
&-&i[T(r)\Delta r\partial_r \tilde a-2\Delta(r)\tilde a]=\tilde bE\,.
\end{eqnarray}
We can thus make the following identification to the continuous model:
\begin{eqnarray}
\label{Tr}
 T(r)&=\frac{\hbar v_F}{\Delta r}\,,\\
\label{Del}
\pm 2\Delta(r)&= \frac{\kappa m+1/2}{r}+\kappa\frac{\pi Br}{\Phi_0}=Q(r)\,.
\end{eqnarray}
The ambiguity introduced by the $\pm$ sign in Eq. (\ref{Del}) can be
settled by looking at the bulk limit of the problem. The choice that
gives the correct answer is $t'(n) = T(r)+ Q(r)/2$ and
$t(n)=T(r)- Q(r)/2$.  If we assume that the wall of the dot is located
at $n=N_l$, then the boundary condition (\ref{zigzag}) is imposed
considering $\tilde\psi^1_{N_l}=0$.  In order to keep the problem
particle-hole symmetric, we consider the case where the effective
chain problem has $N_l$ unit cells.

\begin{figure}[ht]
\begin{center}
\includegraphics*[angle=0,width=8cm]{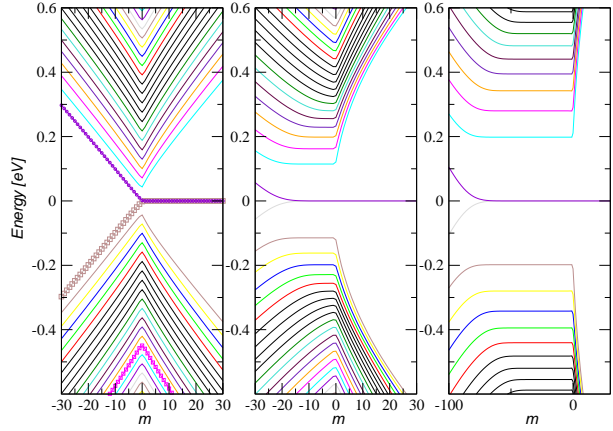}
\caption{(colour on-line) Forty energy levels of a circular graphene
quantum-dot in a finite magnetic field as function of the angular
momentum quantum number $m$. The radius of the dot is $R=70$ nm, three
magnetic fields were used $B=$1, 10, 30 $T$ (from left to right), and
$\kappa=1$. The two surface states (for $m>0$) are represented using
squares and circles.
 \label{fig7}}
\end{center}
\end{figure}

In Fig. \ref{fig7} we represent the numerical solution of
Eqs. (\ref{tb1}) and (\ref{tb2}). It is clear that at small fields the
bands are essentially symmetric for positive and negative $m$ values,
a property that comes from the fact that exchanging $m$ by $-m$ in the
Dirac equation only replaces the role of $\psi^1_n$ and $\psi^2_n$.
With a finite magnetic field the situation changes. Also seen is the
presence of a dispersive edge state. The dispersive part, occurring
for positive $m$, is dependent on the Dirac point. Note the appearance
of the zero energy Landau level upon increasing the magnetic field.
The different behaviour, for large fields, shown by the energy levels
for positive and negative $m$ is associated to the amount of angular
momentum induced by the magnetic field.

\begin{figure}[ht]
\begin{center}
\includegraphics*[angle=0,width=8cm]{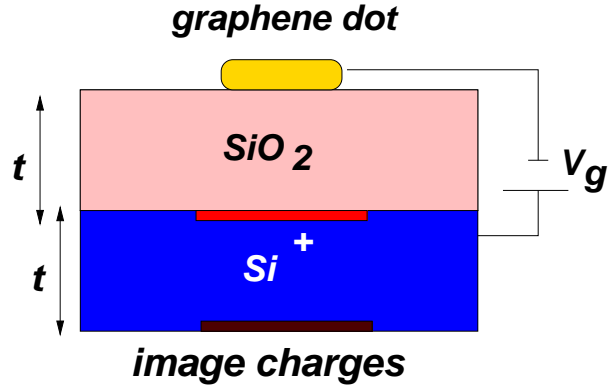}
\caption{(colour on-line)
Lateral view of graphene FET. When the dot is gated the charge accumulates at the
metal-insulator (Si$^+$-SiO$_2$) interface. The thickness of the insulator is $t$
and the applied gate potential is $V_g$. 
 \label{fig8}}
\end{center}
\end{figure}

Let us now discuss how to include the Coulomb interaction in the
calculation. This is important because the screening in the dot may
not be very effective and because the dot may be working under a
regime where it has a net charge density (charged dot). The situation
of a charged dot is represented in Fig. \ref{fig8}. The gate
potential $V_g$ induces either holes or electrons in the dot. This
causes a situation where the dot is charged, since the neutrality case
takes place when the chemical potential is at the Dirac point. If we
take the graphene to be at a potential $V_g$ then the charges
accumulated in the metal-insulator interface have to be at zero
potential. This means that one must use a set of images charges at a
distance $t$ from the interface with exactly the same spatial density
of that formed in the graphene dot.  We then have to describe the
Coulomb interaction of an electron in the dot with both the
self-consistent charge density in the graphene and its image
underneath the metal-insulator interface \cite{brey_hartree}.
The simplest way to include the effect of Coulomb repulsion is by
using the self-consistent Hartree approximation. We analyse here the
effects induced by increasing the number of electrons in the dot using
the Hartree approximation \cite{peeters}. The self-consistent Hartree
potential describes, within a mean field approximation, the screening
of charges within the dot. We assume that a half-filled dot is
neutral, as the ionic charge compensates the electronic charge in the
filled valence band. Away from half filling, the dot is charged. Then,
an electrostatic potential is induced in its interior, and there is an
inhomogeneous distribution of charge. We describe charged dots by
fixing the chemical potential, and obtaining a self-consistent
solution where all electronic states with lower energies than the
Fermi energy are filled. From this calculation we obtain the Hartree
electronic energy bands.  The Hartree approximation should give a
reasonable description when Coulomb blockade effects can be described
as a rigid shift of the electrostatic potential within the dot
\cite{AL91,SET}.  Using the same discretisation procedure as before,
the numerical equations to be solved have now the form
\begin{eqnarray}
\label{Dd1b}
v_H(n) a_n + b_nt(n)+b_{n-1}t'(n-1)&=&a_nE_{j,m}\,, \\
\label{Dd2b}
v_H(n)b_n + a_nt(n)+a_{n+1}t'(n)&=&b_nE_{j,m}\,.
\end{eqnarray}
where the Hartree potential in the continuum, $v_H(r)$, is given by
\begin{eqnarray}
v_H(r)&=&v_0\int r'dr'd\f{\cal K}(\bm r,\bm r', t)
\simeq v_0\frac{R}{N_l}\sum_{n'\neq n}\int^{2\pi}_0d\f\, r'_n\,{\cal K}(\bm r_n,\bm r'_n, t)\,,
\label{vH}
\end{eqnarray}
and
\begin{equation}
{\cal K}(\bm r,\bm r', t)= \frac {\rho(r')}{\sqrt{r^2+(r')^2-2rr'\cos\f}}-\frac {\rho(r')}{\sqrt{r^2+(r')^2-2rr'\cos\f + 4t^2}}
\end{equation}
with the parameter $v_0$ given by $v_0=(e^2/4\pi\epsilon_0\epsilon)$ and
$\rho(r_{n})$ the electronic density at point $r_{n}$,  computed from
\begin{equation}
\rho (r_{n})=gC\sum_m\sum_{j_m\ne j_{m,\rm spur.}}[a_{n,m,j_m}^2+b_{n,m,j_m}^2]/r_n\,,
\label{rho}
\end{equation}
such that the sums over $m$ and $j_m$ are constrained to those energy levels
such that $E_{m,j}\le E_F$ (note that the explicit dependence of $a_n$ and $b_n$ on $m$ and $j_m$
has been introduced in Eq. \ref{rho}), where $E_F$ is the Fermi energy measured
relatively to the Dirac point, $g$ is the spin and valley degeneracy,
and the constant $C$ is given by the normalisation condition on the
disk, $C=(2\pi\Delta r)^{-1}$. The constraint in the $j$ summation
in Eq. (\ref{rho}) is due to the fact that the boundary conditions introduced by the
finite tight-binding chain fails to reproduce accurately the boundary condition
$\tilde\psi_i(r\rightarrow 0)\rightarrow 0$, introducing a spurious mode,
characterised by the quantum number $j_{m,\rm spur.}=N_l+1$ for $m\ge 0$; in order for
sensible results to be obtained these modes have to be removed.

We note that the integral (\ref{vH}) is well behaved since the self-interaction
 has been excluded ($n\neq n'$). Further the $\rho(r_{n'})$ is independent of
$\f$.  The angular integral in Eq. (\ref{vH}) can be formally computed
leading to
\begin{eqnarray}
v_H(n)&=&4v_0\frac{R}{N_l}\sum_{n'\neq n}\left[
\frac {r_{n'}\rho(r_{n'})}{r_n+r_{n'}}\mathbf{K}\left(\frac{4r_nr_{n'}}{(r_n+r_{n'})^2} \right)\right.\nonumber\\
&-&
\left.
\frac {r_{n'}\rho(r_{n'})}{\sqrt{(r_n+r_{n'})^2+4t^2}}\mathbf{K}\left(\frac{4r_nr_{n'}}{(r_n+r_{n'})^2+4t^2} \right)\right]
\,.
\label{vH2}
\end{eqnarray}
with $\mathbf{K}(m)$ defined as
\begin{equation}
\mathbf{K}(m)=\int_{0}^{1}dx[(1-x^{2})(1-mx^{2})]^{-1/2}\,.\label{eq:K}
\end{equation}
The elliptic integral $\mathbf{K}(m)$ can be approximated by an
analytical function \cite{abramowitz}, which reduces the numerical
effort. It is now clear that, due to the Hartree potential, the
problem defined by Eqs. (\ref{Dd1b}) and (\ref{Dd2b}) has to be solved
self-consistently.

One should comment on the fact that for a large dot $R\gg t$ the
contribution from $v_H(\bm r)$ essentially vanishes and the change of
the bands due to the Hartree term is vanishingly small. On the contrary,
for small dots $R\sim t$ and the Hartree renormalisation of the electronic
energy levels can be very important in the case of heavily charged
dots.  

\begin{figure}[ht]
\begin{center}
\includegraphics*[angle=0,width=8cm]{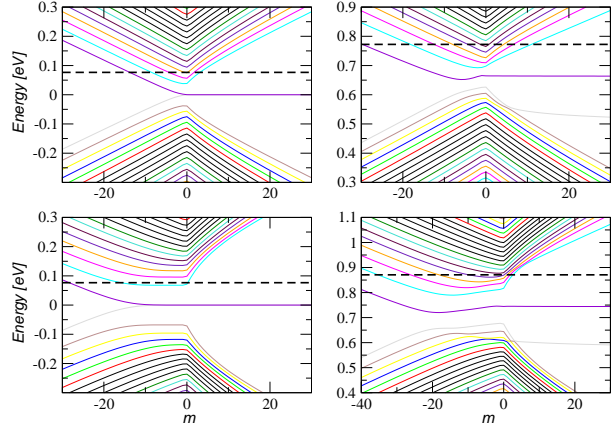}
\caption{(colour on-line) Independent and Hartree energy levels for a
spherical quantum dot with $R=100$ nm, on top of a silicon oxide slab
of $t=100$ nm ($\epsilon=3.9$), for a Fermi energy $E_F=0.077$ eV
(represented by a dashed line).  For $B=1$ T, the number of electrons
in the dot is $N_e=166$ and the magnetic length is $\ell_B=26$ nm; for
$B=3.5$ T, the number of electrons in the dot is $N_e=178$ and the
magnetic length is $\ell_B=14$ nm.  The left panels are the
independent energy bands; the right ones are the Hartree bands.
The top row is for $B=1$ T. The lattice has $N_l=100$.
 \label{fig9}}
\end{center}
\end{figure}

\begin{figure}[ht]
\begin{center}
\includegraphics*[angle=0,width=8cm]{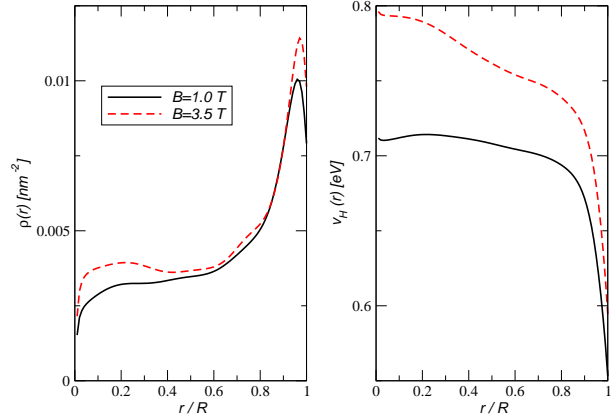}
\caption{(colour on-line) Self-consistent electronic density (left) and
Hartree potential $v_H(n)$ (right) for the same parameters given in
the caption of Fig. \ref{fig9}.
 \label{fig10}}
\end{center}
\end{figure}

In Fig. \ref{fig9} we represent the energy bands of a quantum dot of
radius $R=100$ nm on top of a silicon oxide slab of thickness $t=100$
nm. We used a gate voltage of $V_g=2$ V, which corresponds to a Fermi
energy of $E_F=0.077$ eV for the bulk system, The values of the
magnetic field used were $B=1$ T (top panels) and $B=3.5$ T (bottom
panels).  It is clear that the Hartree bands are renormalised by the
Coulomb interaction.
In Fig. \ref{fig10} we represent the self-consistent density, $\rho(r)$, and Hartree potential,
$v_H(n)$, for two different values of the magnetic field. The
parameters are those given in the caption of Fig. \ref{fig9}. 
The increase in the Hartree potential upon increasing $B$ is due to the increase of the
number of electrons in the dot.

We should comment that in our calculation we have not tried to keep
the number of electrons fixed. This can easily be done, but increases
the computational effort since the chemical potential has to be
self-consistently determined. Instead, we have chosen to keep the Fermi
energy constant, which, of course, leads to a changing in the number
of electrons in the dot with the variation of the magnetic field. Also
the population of the surface states was not included in the
calculation, using a criterion of computational simplicity. In a future
study we shall relax these two constraints.

\begin{figure}[ht]
\begin{center}
\includegraphics*[angle=0,width=8cm]{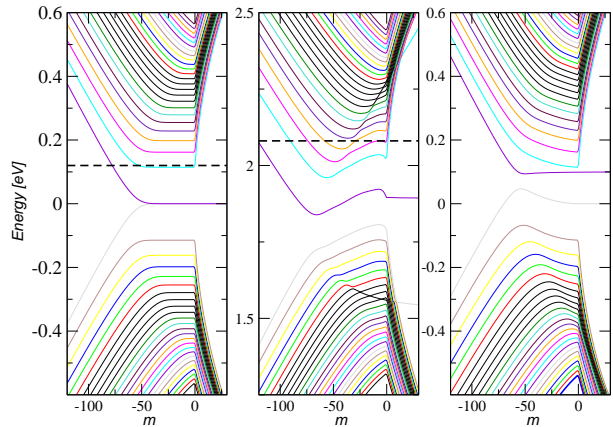}
\caption{(colour on-line) Energy spectrum of a graphene quantum dot at
$B=10$ T. From left to right: independent particle bands, Hartree
bands, and independent particle bands with the confinement potential
(\ref{Vconf}). The parameters are $U_c= 0.1$ eV, $V_g=5$ V, $E_F=0.12$
eV, and the remaining parameters are those used in
Fig. \ref{fig9}. The magnetic length is $\ell_B=8$ nm and the number
of electrons in the dot is $N_e=463$.
 \label{fig11}}
\end{center}
\end{figure}

Another important aspect in quantum dot physics is that of confinement
introduced by the potential creating the dot. The confinement
potential can be either due to etching or to applied gates. In the
case of dots or narrow channels described by the Schr\"odinger
equation, a very popular confinement is that introduced by a parabolic
potential \cite{vanHouten}, since it allows a simple analytical
solution. We choose a confinement potential  given by
\begin{equation}
 V_{conf}(r)=U_c(r/R)^4\,,
\label{Vconf}
\end{equation}
which rises smoothly from the centre of the dot. The prefactor
 $U_c$ is the strength of the potential at the edge of the dot.
As in the case of the Hartree potential, $V_{conf}(r)$ enters in the
diagonal part of the radial Hamiltonian. In Fig. \ref{fig11} we give a
comparison of the energy bands for $B=10$ T. Comparing the left and
the right panels of Fig. \ref{fig11} we see that the Landau levels
become dispersive with $m$ due to the confinement, a result also found
for Landau levels derived from the Schr\"odinger equation
\cite{vanHouten}.  Interestingly, we see that the confinement also
breaks the particle-hole symmetry of the problem, a result found
before for the ribbon problem \cite{peresconf}.

\subsection{The hexagonal and circular dots at the tight-binding level}

In this subsection we want to address the question whether graphene
quantum dots will have or not edge states, starting from the full
solution of the tight-binding Hamiltonian (\ref{Htb}). Edge states in
graphene nanostructures are of particular importance since they can
give rise to magnetism see, e.g., Ref. \cite{Bhowmick}. In order to
access the low energy density of states of dots with physically
relevant sizes (bigger than 10 nm) a Lanczos technique is used, since
the exact diagonalisation of systems of this size becomes
intractable. For a brief introduction to the Lanczos technique, see
e.g. Ref. \cite{White}.

\begin{figure}[ht]
\begin{center}
\includegraphics*[angle=0,width=8cm]{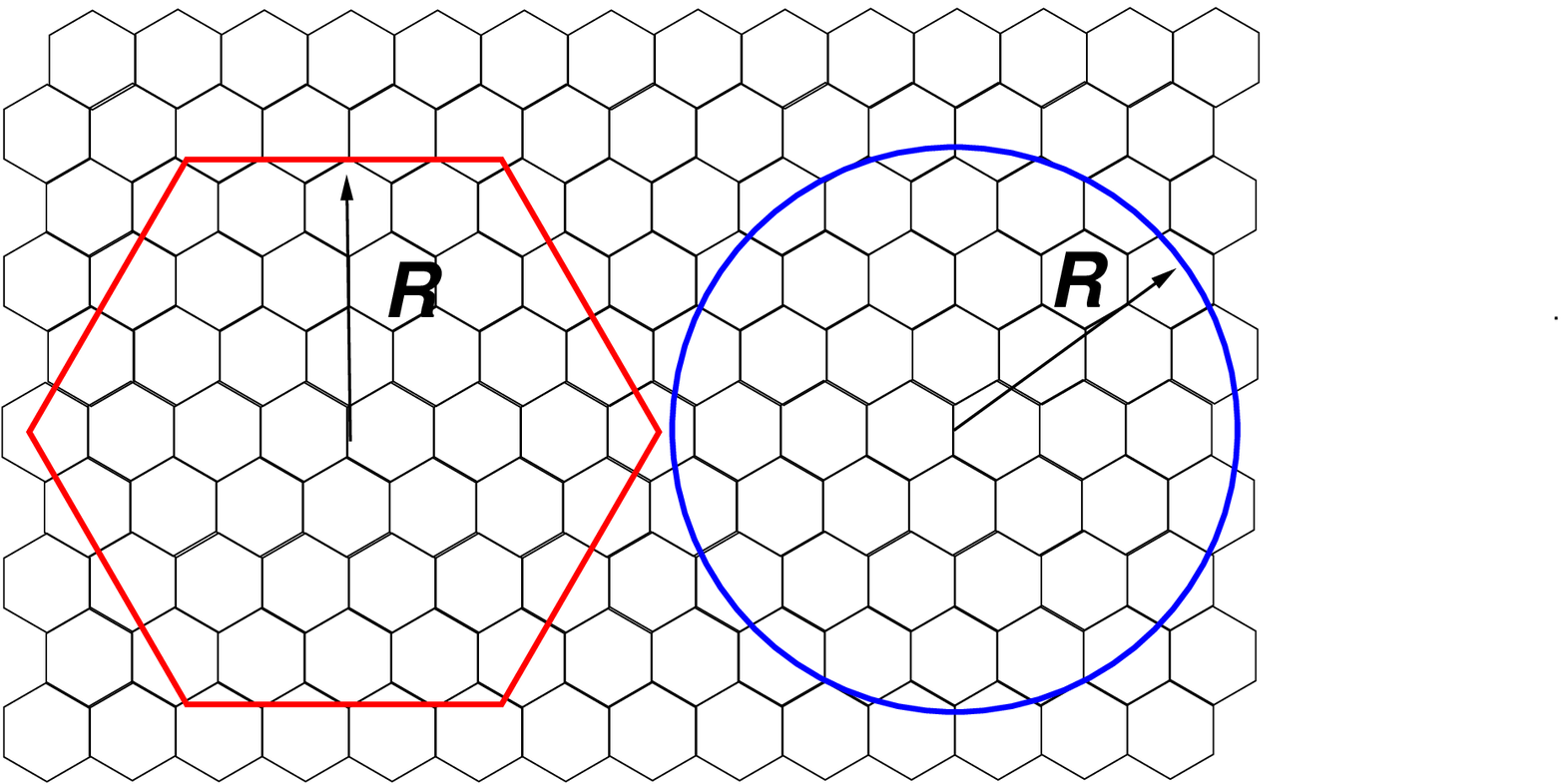}
\caption{(colour on-line)
Representation of a graphene crystallite on top a which an hexagonal or circular
dots can be patterned. The size of the dot is defined by the length $R$.
 \label{Figdots}}
\end{center}
\end{figure}

\begin{figure}[ht]
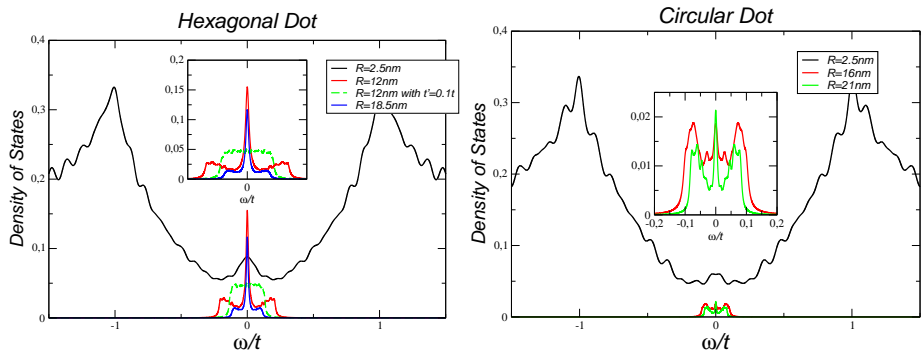

\begin{center}
\includegraphics*[angle=0,width=6cm]{Fig_HexagonalDot2.eps}
\includegraphics*[angle=0,width=6cm]{Fig_CircularDot.eps}
\caption{(colour on-line)
Density of states at low energies for hexagonal (left) and circular dots (right).
The insets are a zoom in of the density of states close to the Dirac point.
Dots of several sizes are represented.
 \label{Figdos}}
\end{center}
\end{figure}

We have chosen to diagonalise dots of {\it circular} and ${\it
hexagonal}$ shape with zigzag-termination, such as those depicted in
Fig. \ref{Figdots}.
Our numerical findings are represented in the Fig. \ref{Figdos}. It is
clear that as the size of the dot grows larger the number of zero
energy states increases, indicating the presence of zero energy-edge
states. When a finite $t'$ is added to the Hamiltonian, the edge
states become dispersive \cite{peres_prb} and there is a reduction of
the density of zero energy states (seen in the hexagonal dot).

\section{Final comments}

In this work a description of the confinement of Dirac electrons in
nano-wires and quantum dots was given. It was shown that, in
principle, it is possible to localise electronic modes in a spatial
region of a nanowire, using a $p-n-p$ gate potential setup. The energy
spectrum of quantum dots in a magnetic field was described taking into
account both the effect of electron-electron interactions, at the
Hartree level, and the effect of confining potentials. The inclusion
of exchange \cite{exch1,exch2,wardana} in this study can in principle
be done. The interesting aspects about this possibility are two-fold:
first, the exchange energy for Dirac electrons is different from that
for the two-dimensional electron gas described by the Schr\"odinger
equation; second, and contrary to the Hartree potential, the exchange
energy of the full electronic system has to be considered, since there
is not an equivalent cancellation effect to that found in the Hartree
potential between the ion background and the valence electrons direct
Coulomb energy.  These aspects will be pursued in a follow-up study
\cite{preparation}.

\ack
NMRP, TS, and JMBLS acknowledge the financial support
from POCI 2010 via project PTDC/FIS/64404/2006 and 
from ESF via INSTANS. The authors thank Daniel Arovas, Ant\^onio Castro Neto, Eduardo Castro, Francisco Guinea,
 and Vitor Pereira for discussions; Andre  Geim is acknowledged for suggestions.

\appendix
\section{Equations for the numerical solution of the scattering problem}
\label{ap0}

Below we give the equations that have been solve numerically when we studied the
scattering problem by a infinite-mass wall. These are:

\begin{itemize}
\item $p=0$:\begin{eqnarray}
\left[\left(1+iz_{0}\right)+r_{0}(1+i\tilde{z}_{0})\right]  +
\frac{2}{i\pi}\left[\left(z_{0}+i\right)+r_{0}(\tilde{z}_{0}+i)\right]  +\nonumber\\
\sum_{m_{\rm odd}>0}\frac{-2}{im\pi}\left[r_{m}(1+i\tilde{z}_{m})\right]  +
\sum_{m_{\rm even}>0}\frac{2}{i(m+1)\pi}\left[r_{m}(\tilde{z}_{m}+i)\right]  =  0\,,
\nonumber\\
\end{eqnarray}
\item $p\neq0$  and even: 
\begin{eqnarray}
r_{p}(1+i\tilde{z}_{p})  +
\frac{2}{i(p+1)\pi}\left[\left(z_{0}+i\right)+r_{0}(\tilde{z}_{0}+i)\right]  +\nonumber\\
\sum_{m_{{\rm odd}}>0}\frac{2}{i(p-m)\pi}\left[r_{m}(1+i\tilde{z}_{m})\right]  +
\sum_{m_{m{\rm even}}>0}\frac{2}{i(p+m+1)\pi}\left[r_{m}(\tilde{z}_{m}+i)\right]  =  0\,,
\nonumber\\
\end{eqnarray}
with the possibility of having $m=p$ in the $m_{\rm even}$ summation
\item $p$ odd 
\begin{eqnarray}
r_{p}(\tilde{z}_{p}+i)  +
-\frac{2}{ip\pi}\left[\left(z_{0}+i\right)+r_{0}(\tilde{z}_{0}+i)\right]  +\nonumber\\
\sum_{m_{{\rm odd}}>0}\frac{-2}{i(p+1+m)\pi}\left[r_{m}(1+i\tilde{z}_{m})\right]  +
\sum_{m_{{\rm even}}>0}\frac{2}{i(-p+m)\pi}\left[r_{m}(\tilde{z}_{m}+i)\right]  =  0\,,\nonumber\\
\end{eqnarray}
\end{itemize}
with the possibility of having $m=p$ in the $m_{\rm odd}$ summation.
It is clear that we can truncate this set of equations to obtain a
set of $N$ equations for the coefficients $r_{0},\dots r_{N-1}$.

\section{General boundary conditions in a quantum dot with infinite mass confinement}
\label{ap1}

In this Appendix we give all the details of how to obtain the boundary
condition of the wave function at the wall of a quantum dot, with the
confinement determined by the infinite mass condition.  We must
compute the wave function in the domain $D$ due to a reflection at the
boundary.  We write the plane-wave inside $D$ as
\begin{equation}
\Psi_{D} = \left ( \begin{array}{ccc} 1 \\ e^{i \theta_{0}} \end{array} \right ) e^{i \vec{k}_{i} \cdot \vec{r}} + R\left ( \begin{array}{ccc} 1 \\ e^{i \theta_{1}} \end{array} \right ) e^{i \vec{k}_{f} \cdot \vec{r}} ,
\label{psiD}
\end{equation}
where from Fig. \ref{fig5} we can conclude that $\theta_{1} = \pi + 2 \alpha - \theta_{0}$,
and $\vec k_i$ and $\vec k_f$ are the momenta of the incident and reflected waves at the boundary of the dot.
In order to calculate $\psi_{2} / \psi_{1}$ (which will make it possible to compute the value of $B$),
we need to discover the value of the reflection coefficient, $R$. We can accomplish this, using the fact required by the Dirac equation, that the components of the spinors must be continuous at the boundary.

First we solve the Dirac equation with a mass $M v_{F}^{2} > E$. For the sake of simplicity, we use the normal and tangential coordinates $n$ and $s$ given by
\begin{eqnarray}
n &=& x \cos \alpha + y \sin \alpha \nonumber \\
s &=& - x \sin \alpha + y \cos \alpha \nonumber
\end{eqnarray}
which implies that,
\begin{eqnarray}
\partial_{x} &=& (\partial_x n)\, \partial_{n} + (\partial_x s)\, \partial_{s} \nonumber \\
\partial_{y} &=& (\partial_y n)\, \partial_{n} +  (\partial_y s)\, \partial_{s} \nonumber \\
\partial_{x} &=& \cos \alpha \partial_{n} - \sin \alpha \partial_{s} \nonumber \\
\partial_{y} &=& \sin \alpha \partial_{n} + \cos \alpha \partial_{s} \nonumber
\end{eqnarray}
resulting in
\[
\partial_{x} \pm i \partial_{y} = (\partial_{n} \pm i \partial_{s}) e^{\pm i \alpha}\,.
\]
Then, for a plane wave in the domain $OD$ (complementary to $D$), $\Psi_{OD} = T \left ( \begin{array}{ll} u \\ v \end{array} \right ) e^{i (k_{n} n + k_{s} s)}$, (where $T$ stands for the transmission coefficient). Solving the Dirac equation explicitly we obtain
\begin{eqnarray}
\Psi_{OD} &=& T  \left ( \begin{array}{ccc} 1 \\ \frac{E - M v_{F}^{2}}{i \hbar v_{F} (q - k)} e^{i \alpha} \end{array} \right ) e^{i k s - q n}\,,
\label{psiOD}
\end{eqnarray}
where we have defined
\begin{eqnarray}
k_{n} &=& i q = i \sqrt{\frac{M^{2} v_{F}^{4} - E^{2}}{\hbar^{2} v_{F}^{2}} + k^{2}} \nonumber
\end{eqnarray}
and chosen the solution that decays for $r \to + \infty$. Further we identified $k_{n} = i q$ and $k_{s} = k$.
Imposing the continuity of the wave functions  (\ref{psiD}) and (\ref{psiOD}) at the boundary of the dot, one obtains
\begin{eqnarray}
1 + R &=& T\,, \nonumber \\ 
e^{i \theta_{0}} + R e^{i \theta_{1}} &=& T i e^{i \alpha}. \nonumber
\end{eqnarray}
Replacing the value of $R$ in Eq. (\ref{psiD}), the wave function in the dot reads
\begin{eqnarray}
\Psi_{D} &=& \frac{1}{\sqrt{2}} \left ( \begin{array}{ccc} 1 \\ e^{i \theta_{0}} \end{array} \right ) e^{i \vec{k}_{i} \cdot \vec{r}} - \frac{1 + i e^{-i (\alpha - \theta_{0})}}{1 - i e^{i (\alpha - \theta_{0})}} \frac{1}{\sqrt{2}} \left ( \begin{array}{ccc} 1 \\ e^{i \theta_{1}} \end{array} \right ) e^{i \vec{k}_{f} \cdot \vec{r}}, \nonumber
\end{eqnarray}
which, after some simple manipulations, allows us to conclude that
\[
\frac{\psi_{2}}{\psi_{1}} = i e^{i \alpha}
\]
and therefore $B = 1$.
\section{The Dirac equation in polar coordinates}
\label{ap2}

To treat problems with circular symmetry, the partial derivatives with
respect to Cartesian coordinates shall be written in polar coordinates
($r,\f$). For the $x$-coordinate, the product rule yields
$\partial_x=(dr/dx)_y\partial_r+(\partial \varphi/\partial
x)_y\partial_\varphi$, where the derivatives are taken for fixed
$y$. The first derivative is obtained using $r=\sqrt{x^2+y^2}$. The
second one uses $\tan\varphi=y/x$ and thus
$(1/\cos^2\f)\partial_\f=-(y/x^2)\partial_x$. This gives
\begin{equation}
\partial_x=\cos\f\partial_r-\frac{\sin\f}{r}\partial_\f\;,
\end{equation}
and analogously
\begin{equation}
\partial_y=\sin\f\partial_r+\frac{\cos\f}{r}\partial_\f\;.
\end{equation}
The Hamiltonian thus reads
\begin{equation}
H_\kappa=-i\hbar v_F 
\left(\begin{array}{cc}
0&e^{-i\kappa\varphi}(\partial_r-\kappa\frac{i}{r}\partial_\varphi)\\
e^{i\kappa\varphi}(\partial_r+\kappa\frac{i}{r}\partial_\varphi)&0
\end{array}\right)\,,
\label{H0}
\end{equation}
where we have introduced the additional quantum number $\kappa$, to account for the two non-equivalent
Dirac points. Let us now define the operators 
\begin{eqnarray}
L_+&\equiv e^{i\varphi}(\partial_r+\frac{i}{r}\partial_\varphi)\,,\\
L_-&\equiv -e^{-i\varphi}(\partial_r-\frac{i}{r}\partial_\varphi)\,,
\end{eqnarray}
which acting on the product of a Bessel function of integer order $m$,
$J_m(kr)$, and a complex exponential, $e^{im\f}$, produce $L_\pm
J_m(kr)e^{im\f}=-kJ_{m\pm1}(kr)e^{i(m\pm1)\f}$. This last result leads
to the construction of the wave function of the free problem in the
form given in Eq. (\ref{psicirc}). In addition, the following
commutators $[L_\f,L_{\pm}]=\pm L_\pm$ and $[L_+,L_-]=0$ allow
us to interpret the operators $L_{\pm}$ as rising and lowering
operators of the angular momentum.

Finally we note that if we consider a ring instead of a disk it is possible to
add a flux through the ring, introducing a vector
 potential $\vec A_\Phi=(\Phi/2\pi r)\vec e_\varphi$. The full Hamiltonian with both a perpendicular magnetic
field and the magnetic flux through the ring 
is  given by
\begin{equation}
H_\kappa=-i\hbar v_F 
\left( \begin{array}{cc}
0&e^{-i\kappa\varphi}(\partial_r-\kappa\frac{1}{r}\left(i\partial_\varphi-\frac{\Phi}{\Phi_0}\right)+\kappa\frac{\pi Br}{\Phi_0})\\
e^{i\kappa\varphi}(\partial_r+\kappa\frac{1}{r}\left(i\partial_\varphi-\frac{\Phi}{\Phi_0}\right)-\kappa\frac{\pi Br}{\Phi_0})&0
\end{array}\right) 
\,,
\label{HwithBandPhi}
\end{equation} 
and its numerical solution can be accommodated within the explained method.


\end{document}